\documentclass[12pt]{article}
\pdfoutput=1
\usepackage{amsmath,amssymb,amsfonts,epsfig,graphicx,euscript}%
\usepackage[pdftex,bookmarks,bookmarksnumbered,linktocpage,pdfstartview=FitH]{hyperref}
\hypersetup{colorlinks,%
citecolor=red,%
filecolor=blue,%
linkcolor=blue,%
urlcolor=blue,%
pdftex}
\usepackage{amsmath}
\usepackage{amsfonts}
\usepackage{graphicx}
\usepackage{caption}
\usepackage{subcaption}
\usepackage{float}
\usepackage[all]{hypcap}
\numberwithin{equation}{section}
\usepackage{cite}
\usepackage{calc}
\newlength{\spacer}
\newsavebox{\mybox}

\newcommand{\bse}{\begin{subequations}}
\newcommand{\ese}{\end{subequations}}
\newcommand{\be}{\begin{equation}}
\newcommand{\ee}{\end{equation}}
\newcommand{\bea}{\begin{eqnarray}}
\newcommand{\eea}{\end{eqnarray}}
\newcommand{\ba}{\begin{array}}
\newcommand{\ea}{\end{array}}

\begin{document}
\hfill
\vspace{0.5cm}
\begin{center}
{ \Large{\textbf{Double Relaxation via AdS/CFT}}} 
\end{center}
\vspace*{0.5cm}
\begin{center}
{\bf S. Amiri-Sharifi$^{1}$, M. Ali-Akbari$^{2}$, A. Kishani-Farahani$^{3}$,\\ N. Shafie$^{4}$}\\%
\vspace*{0.3cm}
{\it {Department of Physics, Shahid Beheshti University G.C., Evin, Tehran 19839, Iran}}\\
{\it {${}^1$$\rm{s}_{-}$amirisharifi@sbu.ac.ir},  {${}^2$$\rm{m}_{-}$aliakbari@sbu.ac.ir},\\ {${}^3$k.farahani$_{-}$a@yahoo.com}, {${}^4$negin.shafie@gmail.com}   }
\end{center}

\begin{center}
\textbf{Abstract}
\end{center}
We exploit the AdS/CFT correspondence to investigate thermalization in an ${\cal{N}}=2$ strongly coupled gauge theory including massless fundamental matter (quark). More precisely, we consider the response of a zero temperature state of the gauge theory under influence of an external electric field which leads to a time-dependent current. The holographic dual of the above set-up is given by introducing a time-dependent electric field on the probe D7-brane embedded in an $AdS_5 \times S^5$ background. In the dual gravity theory an apparent horizon forms on the brane which, according to AdS/CFT dictionary, is the counterpart of the thermalization process in the gauge theory side. We classify different functions for time-dependent electric field and study their effect on the apparent horizon formation. In the case of pulse functions, where the electric field varies from zero to zero, apart from non-equilibrium phase, we observe the formation of two separate apparent horizons on the brane. This means that the state of the gauge theory experiences two different temperature regimes during  its time evolution.

\newpage

\tableofcontents

\section{Introduction and Results}
The AdS/CFT correspondence states that type IIB string theory on $AdS_5\times S^5$ background is dual to ${\cal{N}}=4$  $SU(N)$ super Yang-Mills theory living on the boundary of $AdS_5$ which is a $3+1$-dimensional spacetime \cite{Maldacena, CasalderreySolana:2011us}. In the large-$N$ and large t'Hooft coupling limit the above correspondence reduces to a relation between a strongly coupled gauge theory and a classical gravity. In the mentioned limits, the main and significant property of the correspondence is its strong/weak nature, indicating that one can acquire information about a strongly coupled gauge theory by using its corresponding classical gravity. This duality can be applied at zero temperature as well as at non-zero temperature. As a matter of fact, (thermal) vacuum state in the gauge theory corresponds to the $AdS_5\times S^5$ (AdS-Schwarzschild black hole) background where gauge theory temperature is identified with the Hawking temperature of AdS-Schwarzschild black hole \cite{Witten:1998qj}. Moreover, using the above correspondence, one can study the strongly coupled gauge theories whether in or out of thermal equilibrium. Altogether, the AdS/CFT correspondence has proven itself as a suitable candidate to study strongly coupled gauge theories.

AdS/CFT correspondence, or more generally gauge/gravity duality, has been extensively used to study the issue of quantum quench in strongly coupled gauge theories \cite{Das:2011nk}. Quantum quench is defined as the process of  (rapidly) altering a physical coupling of a quantum system. Since the time dependence of the coupling, both duration and form, is arbitrarily chosen, the system under consideration is driven  out of equilibrium and therefor investigating the evolution of the system becomes an interesting issue. When the coupling changes fast, the so-called fast quench, a universal behavior has been reported for the strongly coupled gauge theories with holographic dual \cite{Das:2014jna}. However, for slow quenches, an adiabatic behavior is observed \cite{Buchel:2014gta}.

The original AdS/CFT correspondence contains only fields in the adjoint representation of the $SU(N)$ gauge group. To describe QCD-like theories, one needs to add matter fields (quarks) in the fundamental representation of the $SU(N)$ gauge group. In the gravity side, this can be done by introducing D-branes in the probe limit into the background \cite{Karch:2002sh}. By probe limit we mean that the back-reaction of the brane on the background is negligible or, in other words, the background we are dealing with remains unchanged\footnote{When the background is $AdS_5\times S^5$(zero temperature), one should worry about the validity of the probe limit in the presence of the external electric field. A simple way to solve this problem is to introduce a small temperature in the background which does not affect the main results. }. If one turns on a constant external electric field on the probe D-brane, matter fields will couple to the electric field and therefore an electric current emerges on the probe brane\footnote{In fact there are two sources that cause an electric current on the probe brane in the presense of an electric field: free charges and Schwinger pair production. In this paper we do not consider free charges and therefore the pair production is the only contributor  to the current. For a more detailed explanation c.f. \cite{Karch:2007pd, Hashimoto:2013mua, Hashimoto:2014dza}.} \cite{Karch:2007pd}. Generalization of the above idea to a time-dependent electric field, resulting in a time-dependent current on the brane, has been studied in \cite{Hashimoto:2013mua}. It is shown that due to energy injection into the system, an apparent horizon forms on the brane at late times, and therefore one can define a time-dependent temperature depending on the location of the apparent horizon.

Electric field quench, i.e. a time-dependent electric field that varies from zero to a constant final non-zero value during a transition time, is considered in \cite{Ali-Akbari:2015gba} at zero temperature and in \cite{Ali-Akbari:2015hoa} at non-zero temperature. It is shown that for fast electric field quenches a universal behavior emerges, indicating that the equilibration time does not depend on the final value of the time-dependent electric field. On the other hand, for slow electric field quenches, the system behaves adiabatically.

Various types of electric field quenches are identified by the form of the functions chosen for the time-dependent electric field. These functions are usually infinitely differentiable ($C^{\infty}$). It would be interesting to study more general functions that don't enjoy this property (i.e. are not $C^{\infty}$). In this paper we investigate the effects of more general functions on the apparent horizon formation on the probe brane. For this reason, first of all, we classify functions into infinite ($C^{\infty}$), half-finite ($C^1$ and $C^2$ ) and finite quenches ($C^1$) in section \ref{different}. Moreover we also introduce pulse functions. Of course, all these functions can be chosen to belong to different differentiability classes. The general characteristic of pulse functions is that they rise up from zero in the finite past and diminish to zero in the finite future, contrary to the quenches where the final value is a non-zero finite value. 

An important point we would like to emphasize on is that after applying a quench, the system under consideration does not reach thermal equilibrium. Instead, it settles down into a non-equilibrium steady state (NESS) with an "effective temperature" but, perhaps, with no well-defined entropy. 
Although our final state is not in thermodynamic equilibrium, and thus thermalization does not indeed take place, the emergence of an effective temperature, which is equivalent to the formation of an apparent horizon in the gravity side, is imprecisely called "thermalization", for example c.f. \cite{Hashimoto:2013mua, Das:2011nk}. Therefore, from now on, by "thermalization" we mean that the flavor fields see an effective temperature, or equivalently, an apparent horizon has formed on the probe brane. 

Our main results can be summarized as follows:
\begin{itemize}
\item Quenches belonging to different $C^r$ classes behave similarly. In other words, system's evolution seems to be independent from the order of differentiability of the quench function. 
\item The slope of the curve describing the apparent horizon formation in $zt$-plane is, physically speaking, meaningful. In fact, when this slope increases indefinitely, the apparent horizon approaches an event horizon on the brane. On the other hand, when the slope decreases, the apparent horizon disappears on the brane in the long time limit.
\item One can define an excitation time, as in section \ref{review}, marking the time at which the system is driven away from equilibrium.  In the fast quench limit, a universal behavior for the excitation time is also observed. 
\item In the case of pulse functions, the system under consideration experiences two out-of-equilibrium regions: first (second), when the electric field increases (decreases) from zero (a constant non-zero value) to a constant value (zero). We show that, after the first energy injection into the system, an apparent horizon with increasing slope forms. Then the second part of the pulse function drives the system out of equilibrium and an apparent horizon with decreasing slope forms. This, of course, is reasonable since the final value of the time-dependent electric field is zero.
\item Even though the EoM is highly non-linear, under certain conditions such as applying a pulse function with a long flat peak, or a double-rise quench with long separation, it behaves linearly.

\end{itemize}


\section{Review of Holographic Electric Field Quench}\label{review}

In this section we present a short review of holographic electric field quench.
This family of quenches can be described as a suddenly applied electric field to a system that consequently induces nontrivial dynamics and may lead to interesting results such as thermalization. In order to analyze these non-equilibrium transitions, the AdS/CFT correspondence for ${\cal{N}} = 2$ supersymmetric QCD-like, at zero temperature, has been used.  This time-dependent phenomena is described in the gravity theory by the dynamics of a D7-brane, supersymmetrically embedded in an $AdS_5 \times S^5$ spacetime, subject to a dynamical electric field.

We start up with an $AdS_5 \times S^5$ background, with the following metric:
\be\label{metric}
ds^2=\frac{R^2}{z^2}(-dt^2+dz^2+d\vec{x}^2)+R^2d\Omega_5^2,
\ee
here $d\vec{x}^2=dx_1^2+dx_2^2+dx_3^2$ and the radial coordinate is denoted by $z$. The boundary of the above background is located at $z=0$ where, according to the AdS/CFT correspondence, the gauge theory lives. We also split the metric on the five-dimensional sphere $S^5$ as
\be
d\Omega_5^2 = d\theta^2 + \cos^2\theta d\psi^2 + \sin^2\theta d\Omega_3^2.
\ee
It is well-known that adding a $D7$-brane at the probe limit to this background is equivalent to introducing fundamental matter (quark) to $\cal{N}$ $= 4$ SYM theory where the resulting gauge theory preserves $\cal{N}$ $ = 2$ supersymmetry \cite{Karch:2002sh}. At low energies, the dynamics of a D7-brane is described by the Dirac-Born-Infeld (DBI) action
\be\label{action}
S_{DBI}=-\tau_{7}\int d^8\xi \ \sqrt{-det(g_{ab}+(2\pi\alpha')F_{ab})}, 
\ee
where $\tau_7$ is the brane's tension, $g_{ab}=G_{MN}\partial_aX^M\partial_bX^N$ is the induced metric on the brane, and $G_{MN}$ is the metric of the ten-dimensional $AdS_5 \times S^5$ space introduced in \eqref{metric}. Also, $X^M$ is a function of $\xi^a$ that traces the brane's position, and $F_{ab}$ is the field strength of the gauge field living on the brane. Moreover, $\alpha'=l_s^2$ ($l_s$ is the string length) and $\xi^a$ runs over all eight free dimensions of the D7-brane, i.e. $\xi^a = (t,x_1,x_2,x_3,z,\omega_1,\omega_2,\omega_3)$, where $(\omega_1,\omega_2,\omega_3)$ is the coordinates on the unit 3-sphere $S^3$  with $d\Omega_3^2$ metric.

Switching on a non-dynamical electric field $E(t)=E_0$ on the brane would result in a constant electric current $j_0=(2\pi\alpha'E_0)^{3/2}$ \cite{Karch:2007pd}. While smooth transition of the electric field $E(t)$, between zero and a final value of $E_0$, would generate a time-dependent current $j(t)$ that relaxes to the equilibrated current $j_0$, in the distant future \cite{Hashimoto:2013mua, Ali-Akbari:2015gba}. In order to calculate the time-dependent current let us start with the following embedding for the D7-brane 
\be 
\begin{array}{ccccccccc}
   & t      & x_1    & x_2    & x_3    & z      & \Omega_3 & \theta & \psi,   \\
D7 & \times & \times & \times & \times & \times & \times   &      &      ,  \label{embedding}
\end{array}
\ee 
where $\theta$ and $\psi$ are, in general, functions of the eight free parameters $\xi^a$, as above. We also, adopt an ansatz for the gauge field as below ($x \equiv x_1$),
\be
A_x=-\int_{t_0}^tE(s) ds + h(t,z), \label{ansatz}
\ee
and all other components of the gauge field are set to zero \footnote{One can check that the above ansatz for the gauge field is consistent with the equations of motion for $A_a$ and a time-dependent electric field does not necessarily produce a magnetic field.}. Here $t_0$ is an initial time mark that can be set, for instance, equal to $-\infty$. Moreover, we set $\theta = 0$ and $\psi = 0$, which correspond to massless fundamental matter in the gauge theory, for example c.f. \cite{Hoyos:2011us}. It is easily checked that the above ansatz is compatible with the equations of motion for $\theta$ and $\psi$ fields obtained from \eqref{action}, and this solution has the smallest on-shell action corresponding to the lowest energy in the field theory.

In \eqref{ansatz}, $h(t,z)$ is a dynamical degree of freedom and $E(t)$ is the external electric field, which is the main source of manipulation of the system. As mentioned before, in case of a constant electric field, one would expect to observe a constant output current. On the other hand in the more interesting case of a time-dependent electric field that rises up from zero to a constant value of say $E_0$, called a quench, it is expected that the system would go through a non-equilibrium phase and only then, asymptotically approach the same current as in the case of a constant electric field. We shall restrict the time dependency of the electric field $E(t)$ to the standard quenches, and their combinations, discussed below.

Using the embedding \eqref{embedding} and the ansatz \eqref{ansatz}, the DBI action \eqref{action} for the massless fundamental matter (assuming $R=1$ and replacing $2\pi\alpha'E\equiv E$) reduces to
\be
\mathcal{L} \propto \frac{1}{z^5} \sqrt{w} ,
\ee
where $w = 1-z^4 \left[(\partial_tA_x)^2-(\partial_zA_x)^2\right]$.
Consequently, the equation of motion for the only non-vanishing gauge field, i.e. $A_x$, is as follows 
\be
\partial_z \left(\frac{\partial_zA_x}{z\sqrt{w}} \right)-\partial_t \left(\frac{\partial_tA_x}{z\sqrt{w}} \right)=0.
\label{eom}
\ee
The above second order partial differential equation \eqref{eom} requires a suitable set of initial-boundary conditions which is
\be
\begin{aligned}
h(t,z\rightarrow 0)&=0,\quad \partial_t h(t,z\rightarrow 0)=0,\\
h(t\rightarrow t_0,z)&=0, \quad \partial_z h(t\rightarrow t_0 ,z)=0.
\end{aligned}
\ee

After solving the equation of motion for the dynamical filed $h(t,z)$, the current is given by the second derivative of $h(t,z)$ with respect to $z$ at the boundary  \cite{Karch:2002sh}
\be
\partial_z^2h(t,z\rightarrow0)=\left\langle J^x(t) \right\rangle /\cal{N},
\ee
where $\cal{N}$ is a constant and $\left\langle J^x(t) \right\rangle$, or simply $j(t)$ henceforth, is the time-dependent output current.

With the time-dependent current at hand, besides the late-time behavior of the current, $j(t\rightarrow +\infty)$, we are also interested in the non-equilibrium part in the middle. It is convenient to define two important physical time-scales of the system, i.e. equilibration time, $t_{eq}$, and excitation time, $t_{exc}$, as follows
\begin{itemize}
\item {\bf{Equilibration Time ($t_{eq}$)}:}
is the time when $j(t)$ falls within a relative certain percentage (typically $\%5$ ) of the final equilibrated value $j_0$ and stays blew this limit afterwards. In other words, $t_{eq}$ is the smallest positive value for which $t_{eq}<t$ implies $\mid\frac{j(t)-j_0}{j_0}\mid < 0.05$
\item {\bf{Excitation Time ($t_{exc}$)}:}
is the time when $j(t)$ rises above, say, $\%5$ mark of the current's maximum. Meaning that, if $j_{max}$ is the maximum value for $j(t)$, then $t_{exc}$ is the smallest positive solution for $j(t)=0.05 j_{max}$. In other words, for $t\in[0,t_{exc})$ we have  $\mid\frac{j(t)-j_{max}}{j_{max}}\mid < 0.05$.
\end{itemize}

Also, of extreme interest, is the time when apparent horizon is formed on the probe brane. Apparent horizon formation is an interestingly unique procedure that belongs at the heart of AdS/CFT interpretation of thermalization \cite{Hashimoto:2013mua, Das:2011nk, AliAkbari:2012hb}.
In the static massless case, it is shown in \cite{Hashimoto:2013mua} that for constant electric fields, the effective metric \cite{Seiberg:1999vs}, due to gauge fluctuations, induces a metric on the D7-brane that exhibits a black hole singularity at $z_p=E_0^{-1/2}$. The temperature corresponding to this black hole is given by 
\be\label{temperature static}
T = \sqrt{\frac{3}{8\pi^2}} \frac{1}{z_p} = \sqrt{\frac{3}{8\pi^2}}E_0^{1/2}.
\ee

In the dynamical massless case, an apparent horizon is formed at certain $z=z_{AH}(t)$, which asymptotically approaches $z_p$ for sufficiently large $t$. The time dependency of $z_{AH}(t)$, which is of special interest, comes straight from the definition of apparent horizon. Apparent horizon is defined as a hypersurface whose volume does not change when expanding along an outward null geodesic. If $\tilde{G}_{ab}$ is the effective metric which the degrees of freedom on the D7-brane feel \cite{Seiberg:1999vs}, then the volume $V_{AH}$ of a hypersurface for a fixed $t$ and fixed radial $z$ is \cite{Hashimoto:2013mua}
\be
V_{AH} = \int d^3 x^i d^3 \theta^I \sqrt{\displaystyle \prod_{i=1}^{3} \tilde{G}_{ii} \prod_{I=1}^{3} \tilde{G}_{II} },
\ee
(c.f. Appendix B in \cite{Hashimoto:2013mua}), which, for our case equals 
\be
V_{AH} = V_3 . V(S^3). \frac{\tau_7^{-3/2}}{z^3}\lbrace1+z^4(F_{1z}^2-F_{t1}^2)\rbrace^{3/4}\lbrace1+z^4(F_{1z}^2-F_{t1}^2)\rbrace^{1/2}.
\ee
Here, $V_3$ is the volume of $(x^1,x^2,x^3)$ space, and $V(S^3)$
is the volume of the unit 3-dim sphere ($=2\pi^2$). A null vector in the $(t,z)$ sector, which is obviously orthogonal to the aforementioned hypersurface, satisfies $G_{ab}v^av^b=0$. Using this vector, the equation for the apparent horizon reads \cite{AliAkbari:2012hb, Hashimoto:2013mua}
\be\label{ah}
\delta V_{AH}(t,z_{AH}(t)) \equiv (v_t\partial_t+v_z\partial_z)V_{AH}(t,z)\mid_{z=z_{AH}(t)}=0.
\ee
Notice that, due to linearity of \eqref{ah} in $v_t$ and $v_z$, only the ratio $v_t/v_z$ is needed which can be obtained from the nullity equation for $v^\mu$.

\section{Quench and Pulse}\label{different}
In this section, we introduce and classify different functions for time-dependent electric field that, along with their appropriate combinations, will be used as one of the boundary conditions for the equation of motion \eqref{eom}. The reason to apply a certain function in a special situation, is discussed casewise.

\subsection{Standard Quenches}
Here, we provide a comprehensive account of quench functions that will be used throughout this paper. By the term quench we mean turning on the electric field smoothly from zero to a final maximum amount of $E_0$. Even though the form of the quench function will prove irrelevant in the final account (like final temperature), when we derive physical conclusions regarding the system under study, there are some subtleties that are worth mentioning.

\begin{figure}
\begin{center}
    \includegraphics[width=6.5 cm]{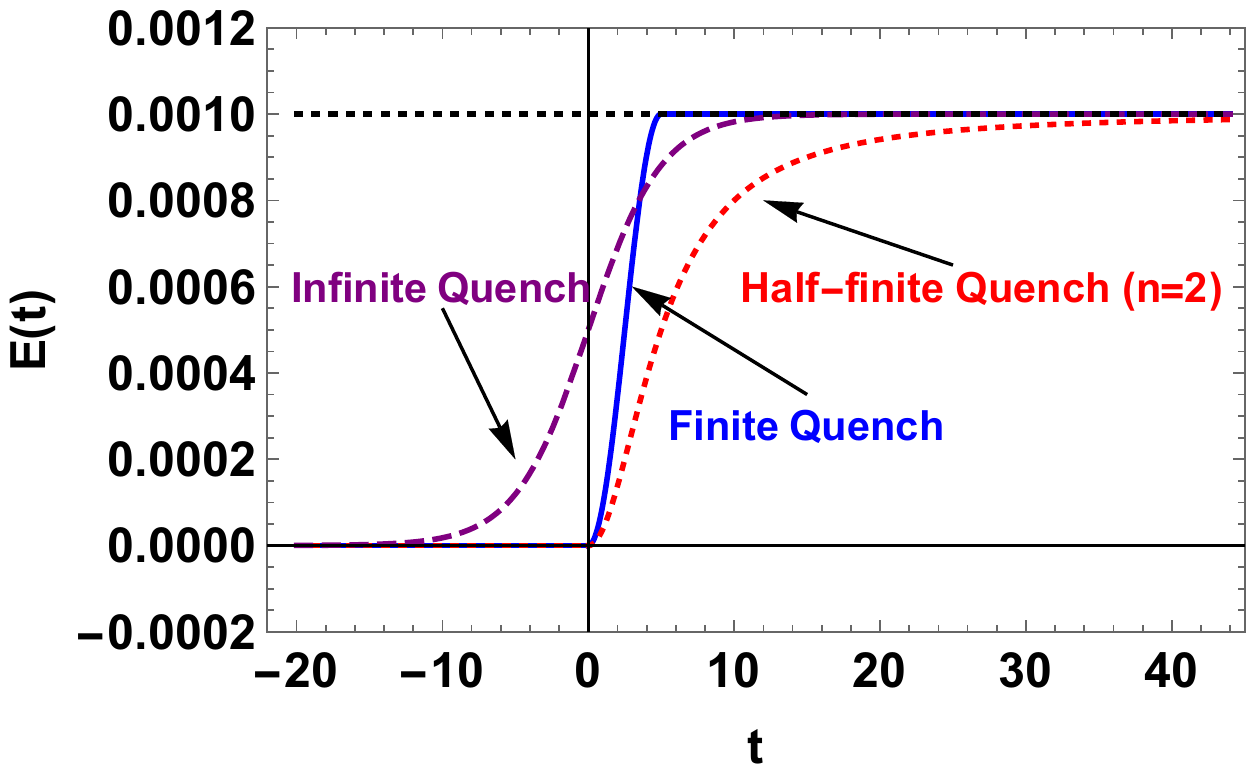}
    \includegraphics[width=6.5 cm]{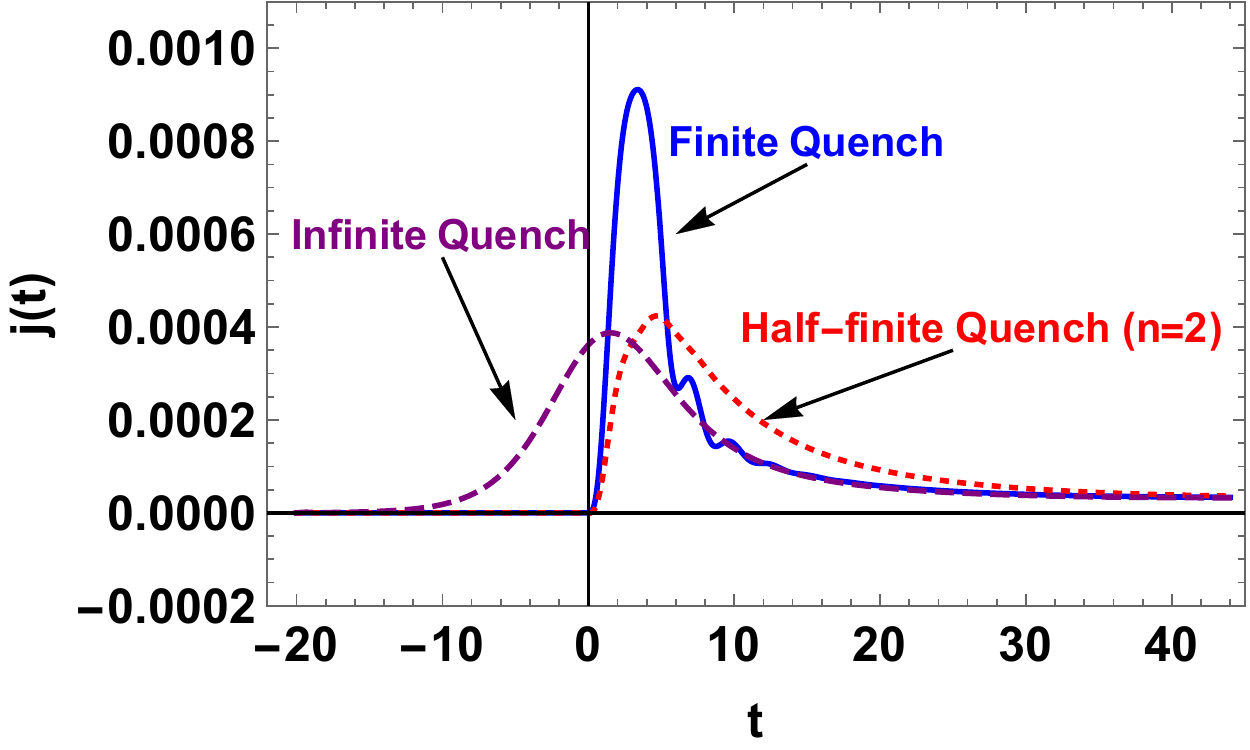}
  \caption{All three standard quench functions (left), and their resulting currents (right), drawn for $E_0=10^{-3}$ and $k=5$.}{\label{allquenches}}
  \end{center}
\end{figure} 

\begin{itemize}
\item {\bf{Infinite Quench:}} In this type of quench functions, the electric field rises up from zero at very distant past, $-\infty$, and reaches its final value $E_0$ at distant future, $+\infty$. A typical function of this class is
\be\label{finite1}
E_{IF}(t)=\frac{E_0}{2}\left(1+\tanh(\frac{t}{k})\right).
\ee
Here $k$ presents a (temporal) measure of elevation of the electric field and one would expect to, practically, reach extreme values $E_0$ and zero, within a few steps of length $k$ in the positive or negative directions, respectively. This function is smooth ( and in fact $C^\infty$) and supposedly should work fine as boundary condition for our second order partial differential equation \eqref{eom}. On the con side, the process of turning on the electric field and reaching final value of $E_0$, mathematically speaking, is meaningless in this case, as $E=0$ and $E=E_0$ happen at $\pm\infty$.

\item {\bf{Half-finite Quench:}} In this type, the electric field remains constant until zero, and then suddenly but rather smoothly, rises up to its final value $E_0$ asymptotically. The process of turning on, happens exactly at $t=0$ but the final value $E_0$ is reached only after an infinite amount of time. These functions are usually presented by a piecewise formula and are specially devised such that the joint at $t=0$ is at least $C^1$ so that the necessary requirement for a second order partial differential equation such as \eqref{eom} is satisfied. A typical function of this type is
\be\label{half-finite}
E_{HF}(t) = E_0
  \begin{cases}
   0 & \text{if } t \leq 0, \\
   1-\displaystyle{\frac{1}{1+(\frac{t}{k})^n}}       & \text{if } t > 0.
  \end{cases}
\ee
Here $k$ plays the same role as in the previous case and is intuitively, directly proportional to the time needed to reach maximum value of $E_0$. Note that for sufficiently large $n$, the continuity of the function and its higher derivatives, is guaranteed. Interestingly, our numerical results \footnote{All numerical calculations in this paper are done using WOLFRAM MATHEMATICA software.} suggest that these functions behave better as $n$ increases. 
 
\item {\bf{Finite Quench:}} In this category, the electric field is turned on exactly at $t=0$ and reaches its exact final maximum value of $E_0$ at some finite time. A typical function of this type which is widely used in the literature is
\be\label{finite}
E_{F}(t) = E_0
  \begin{cases}
   0 & \text{if } t \leq 0, \\
   \frac{1}{2}(1-\cos(\frac{\pi t}{k}))       & \text{if } 0<t<k, \\
   1 & \text{if } t \geq k,
  \end{cases}
\ee
again, $k$ plays the same role as before. (Even though this function is only $C^1$, it has been successfully used in this paper and elsewhere).
\end{itemize}

For comparison, we have solved the equation of motion \eqref{eom} for all three quench types and the resulting currents are portrayed in figure \ref{allquenches}. There's a visible difference in the height of the peaks of the three currents. This is basically due to the fact that these quenches behave differently at $t \sim 0$, the bigger the shock to the system, the higher the peak. Also, by increasing $n$ in the half-finite type, this transition becomes increasingly smoother, so that the peak moves lower.

There are, also, numerous "spline" functions in this category that are of polynomial type which basically resemble $E_F(t)$. As an example we mention the following fifth degree function, which is a $C^2$ function, unlike $E_F$ that is only $C^1$.
\be
E_{S}(t) = E_0
  \begin{cases}
   0 & \text{if } t \leq 0, \\
   6(\frac{t}{k})^5-15(\frac{t}{k})^4+10(\frac{t}{k})^3   & \text{if } 0<t<k, \\
   1 & \text{if } t \geq k.
  \end{cases}
\ee
This function and some other similar to it have been studied in \cite{Ali-Akbari:2015gba}.
\subsection{Opposite Quench}
\begin{figure}
\begin{center}
    \includegraphics[width=6.5 cm]{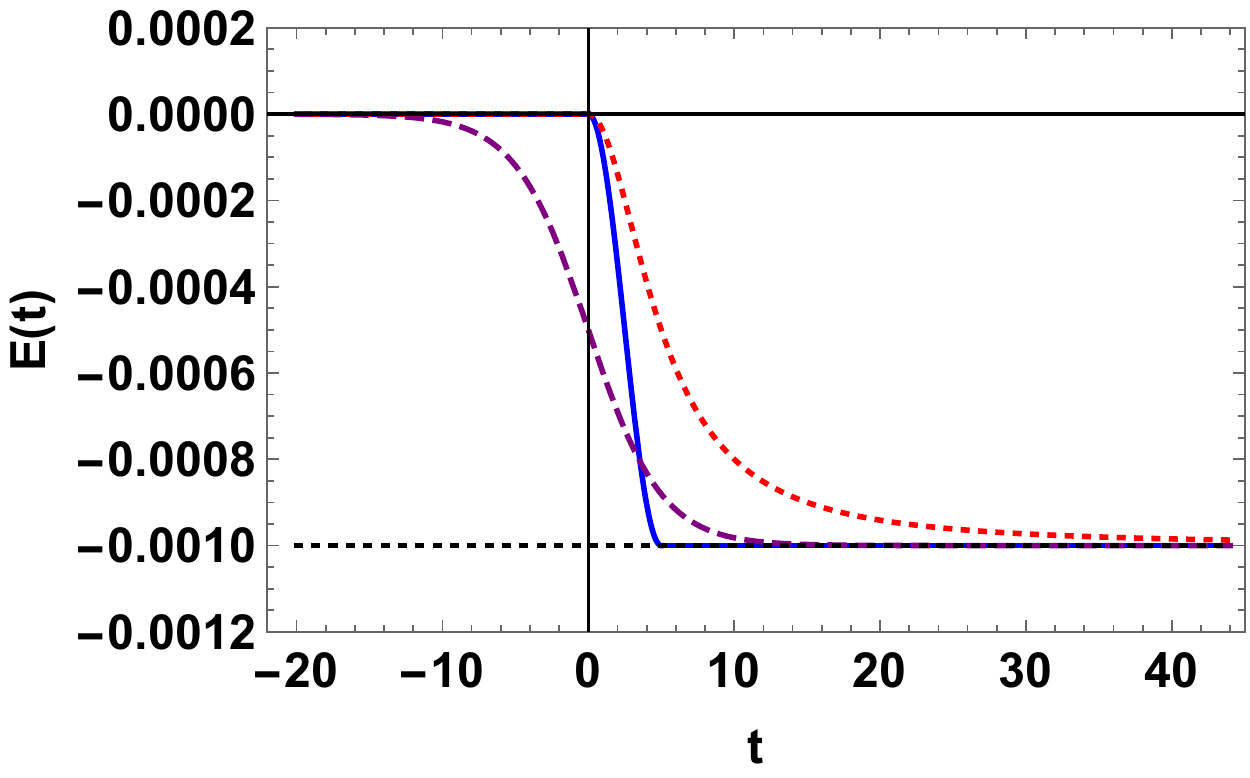}
      \includegraphics[width=6.5 cm]{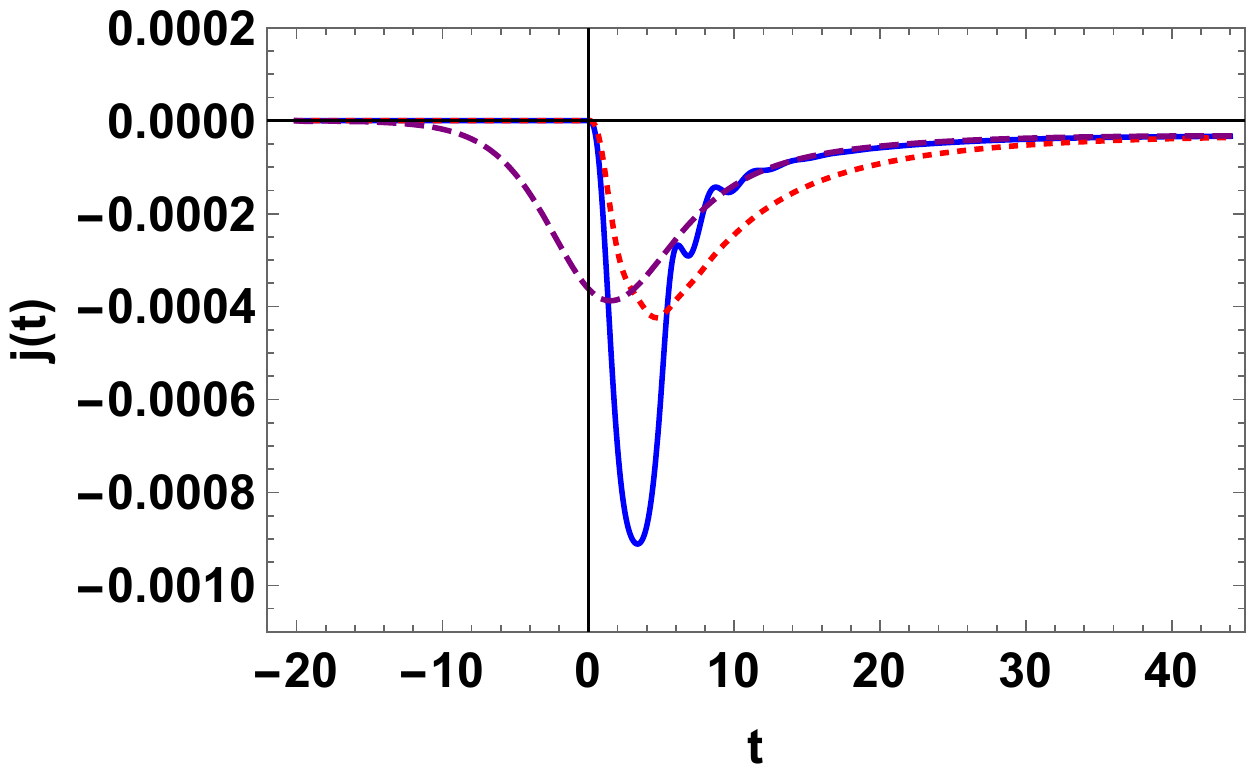}
  \caption{Standard opposite quench functions (left), and their resulting currents (right), drawn for $E_0=10^{-3}$ and $k=5$.}{\label{opposite}}
\end{center}
\end{figure}

Another obvious, yet interesting, quench type arises from considering an opposite (in sign) of the standard quenches. This would mean feeding $-E(t)$ (where $E(t)$ is a standard quench) to the system. As it could be easily seen from \eqref{eom}, this procedure results in the same current with an extra minus sign. The late time behavior for $t \rightarrow + \infty$, and also for $t \rightarrow - \infty$, are the same as the standard quench, with a subtle change of sign in the final value for the current. This is clearly shown in figure \ref{opposite}.

\subsection{Pulse}
\begin{figure}
\begin{center}
    \includegraphics[width=6.5 cm]{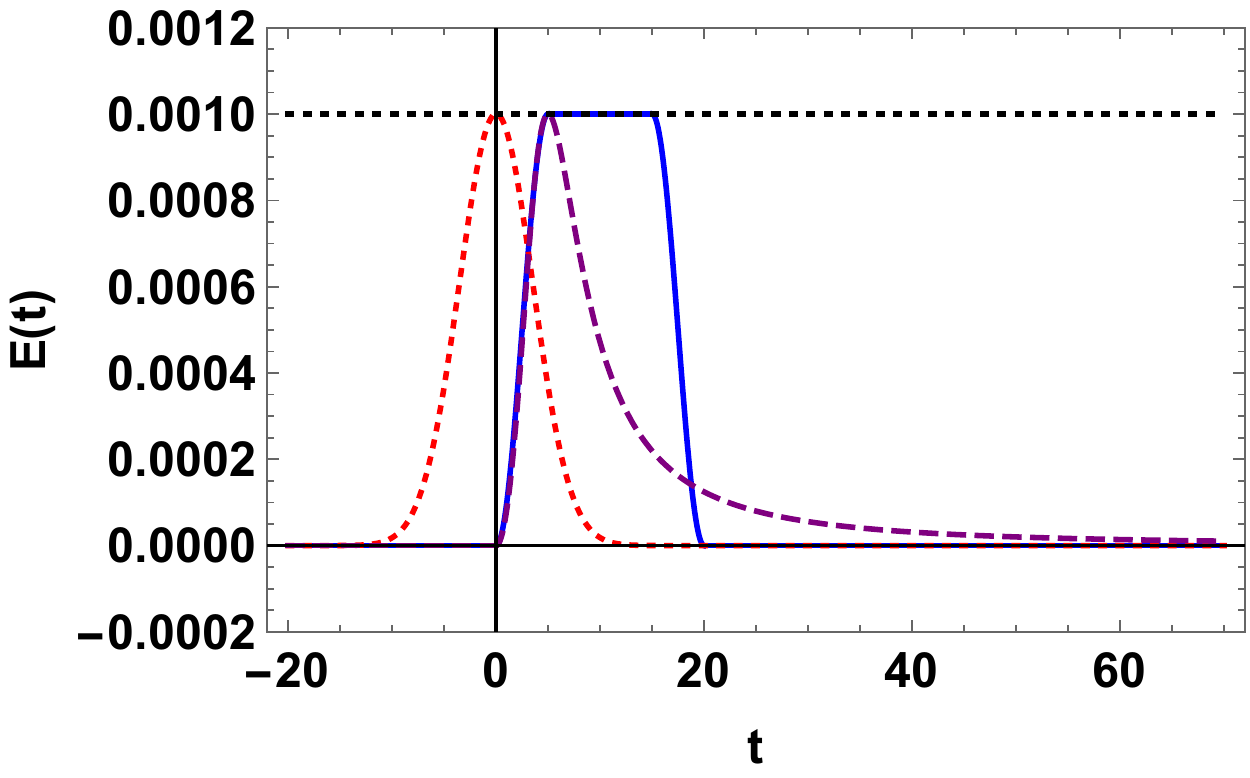}
    \includegraphics[width=6.5 cm]{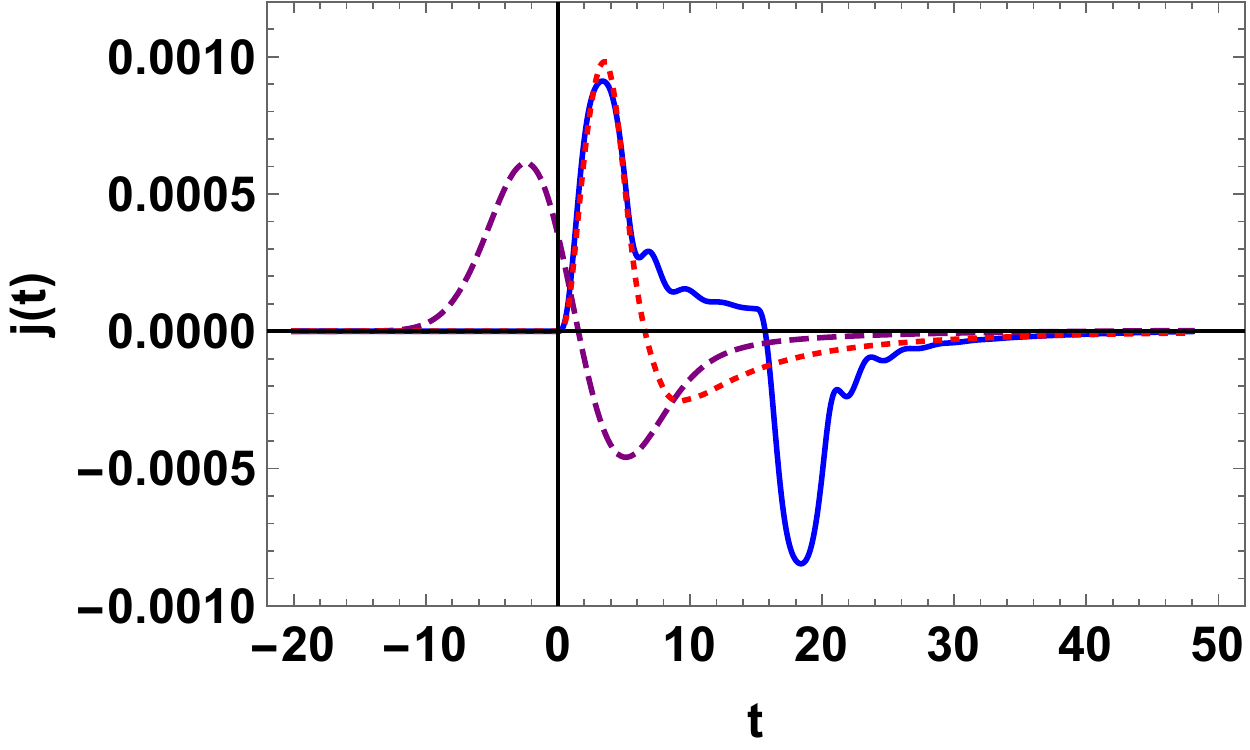}
  \caption{Pulse functions and their resulting currents (drawn for $E_0=10^{-3}$, $k=5$ and $l=10$). Blue, Purple (dashed), and Red (dotted) curves show finite, infinite, and half-finite type quenches and their currents, respectively.}{\label{pulses}}
\end{center}
\end{figure}

This very specific type, presents a smooth transition in the form of zero to zero at the electric field level. In this type, the electric field is zero at the late times ($t\rightarrow\pm\infty$), but it is non-vanishing in the middle ($t\sim0$). There exist also, infinite, half-finite and finite forms of this type, whose functionalities are given by
\bse\begin{align}
E_{I}(t)&=e^{-(t/k)^2},\\
E_{HF}(t)&=
 \begin{cases}
   0 & \text{if } t \leq 0, \\
   \displaystyle{\left(\frac{(n-1)^{1-\frac{1}{n}}}{n}\right)\frac{(\frac{t}{k})^{n-1}}{1+(\frac{t}{k})^n}}       & \text{if } t > 0,
  \end{cases}\\
E_{F}&=
 \begin{cases}
   0 & \text{if } t \leq 0, \\
   \frac{1}{2} \left(1-\cos(\frac{\pi t}{k})\right) & \text{if } 0 \leq t \leq k,\\
   1 & \text{if } k \leq t \leq k+l, \\
   \frac{1}{2} \left(1+\cos(\frac{\pi(t-l)}{k})\right) & \text{if } k+l \leq t \leq 2k+l, \\
   0 & \text{if } 2k+l\leq t. 
  \end{cases}\label{pulsefunctions}
\end{align}\ese
These pulses are illustrated in figure \ref{pulses}, along with their resulting currents. It is clearly seen that applying such a pulse to the system will disturb it for a while (depending on the magnitude of $k$ and $l$), but the system will eventually return to the same original equilibrium state.
An interesting feature of the pulse-currents is the appearance of double "bumps", one in the positive $j(t)$ direction and the other in the negative $j(t)$ direction. These bumps appear as a result of severe changes in the electric field and coincide with the processes of turning the electric field on and off. The most intriguing aspect of this phenomenon will be discussed when apparent horizon formation is studied.

\section{Excitation Time}
\begin{figure}
\begin{center}
    \includegraphics[width=6 cm]{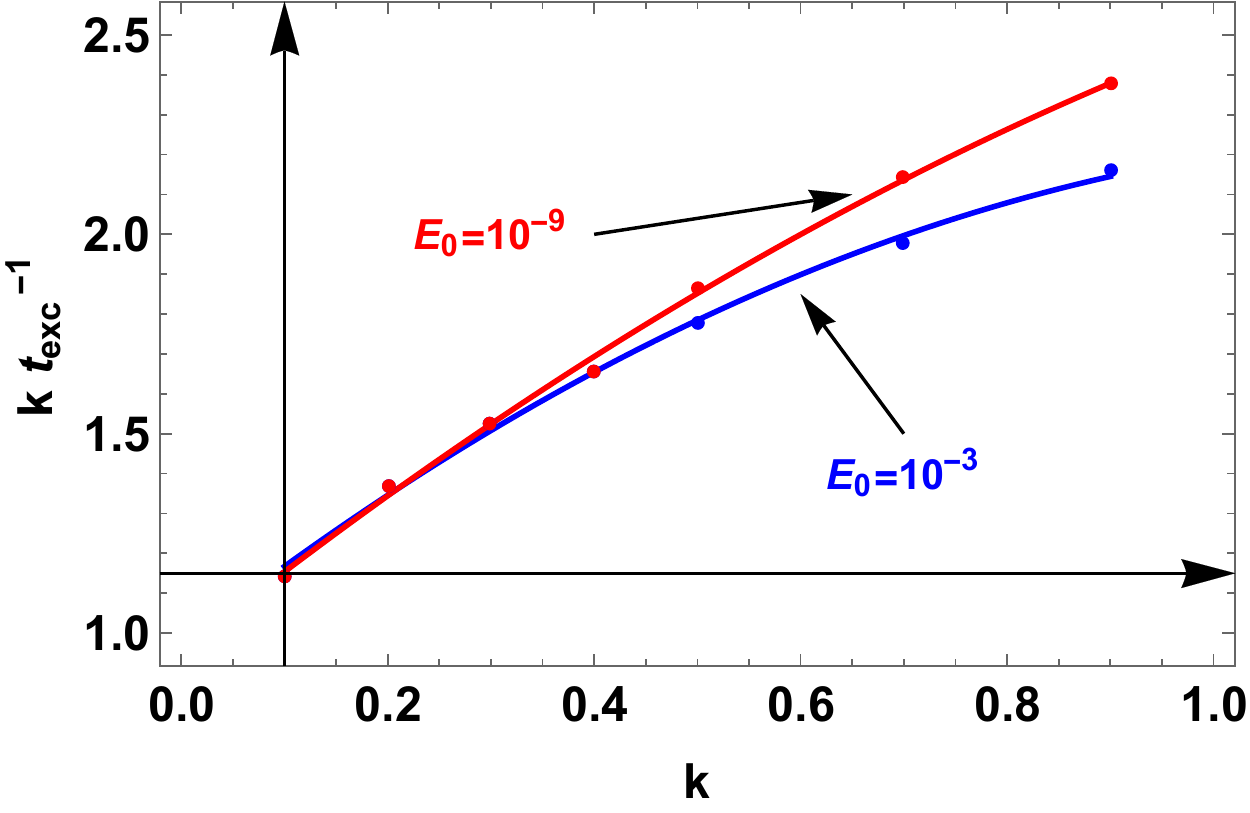}
    \includegraphics[width=6 cm]{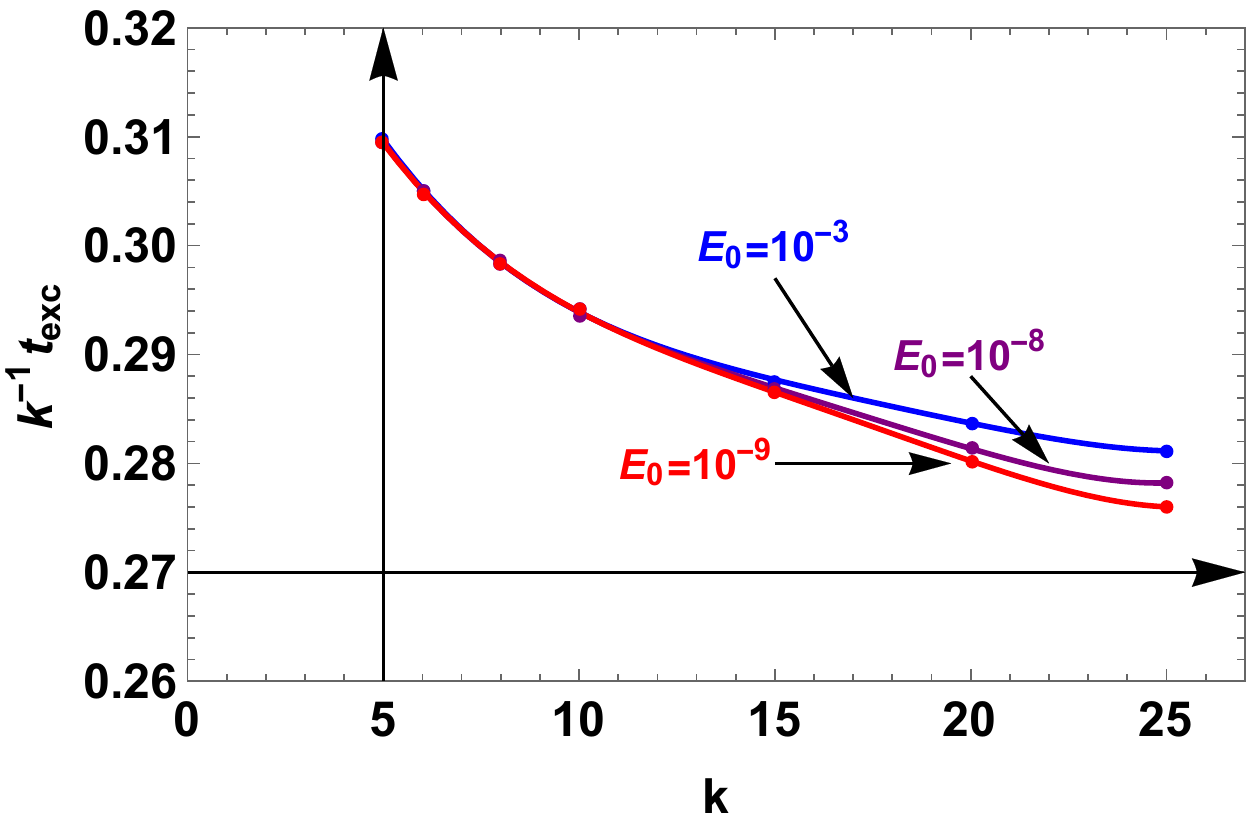}
  \caption{Re-scaled excitation time and its inverse versus $k$, for half-finite function \eqref{half-finite} with $n=2$. Left panel: fast quench and right panel: slow quench.}{\label{excitation}}
\end{center}
\end{figure}
The behavior of the re-scaled equilibration time $k^{-1}t_{eq}$ has already been discussed (for finite type quenches), c.f. \cite{Ali-Akbari:2015gba}. 
Indeed, there it is shown that for slow quenches the system under study behaves adiabatically. Moreover, a universal behavior is reported for the fast quench limit, meaning that the inverse value of the re-scaled equilibration time, i.e. $k t_{eq}^{-1}$, is independent of the electric field's final value. Similarly, in order to study the system's behavior during the process of turning the electric field on, the excitation time, as defined in section \ref{review}, is investigated for fast and slow quenches, corresponding to small and large $k$, respectively.
To convey a schematic picture of the excitation phenomena, figure \ref{excitation} is presented which illustrates $t_{exc}$ versus $k$ for fast and slow quenches of half-finite type.
Fast quenches are depicted for $k=0.1-0.9$ and slow quenches are depicted for $k=5-25$. The re-scaled excitation time and its inverse, both dimensionless, are respectively introduced as $k^{-1}t_{exc}$ and $kt_{exc}^{-1}$. 

\begin{figure}
\begin{center}
    \includegraphics[width=6 cm]{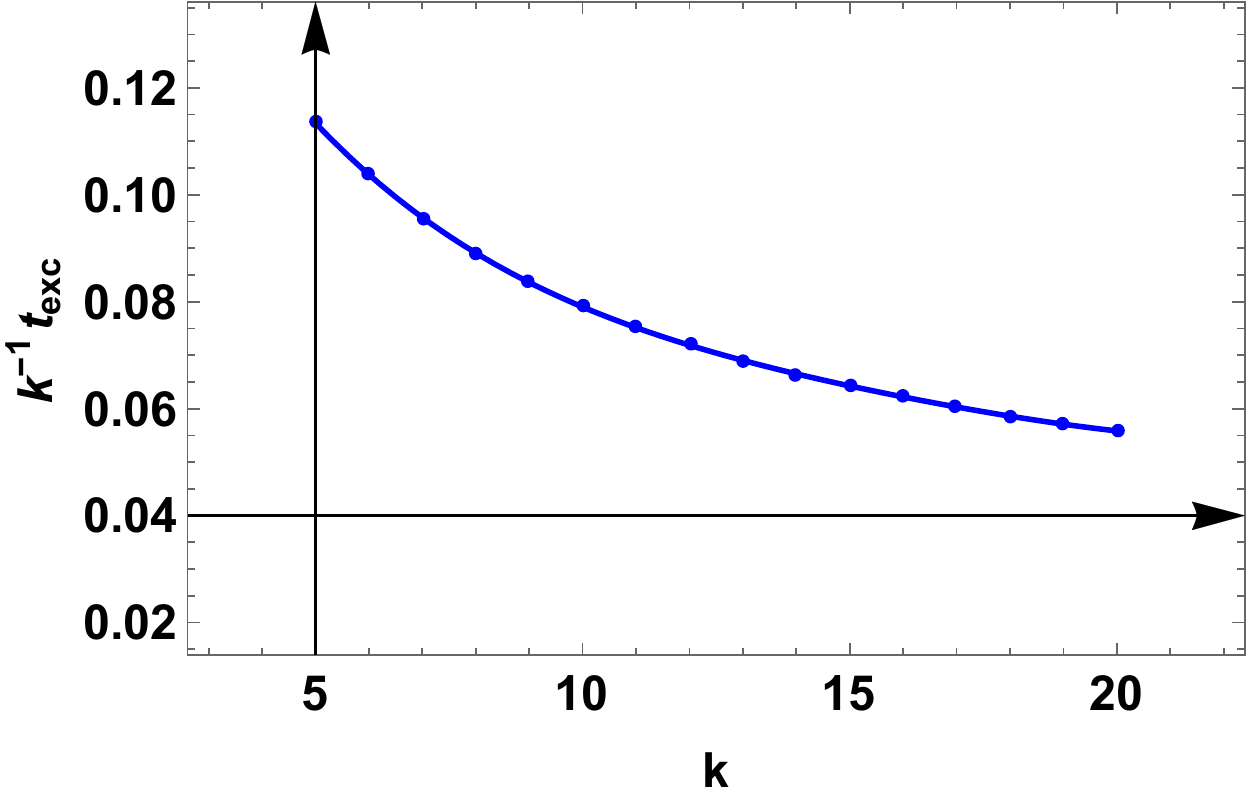}
    \includegraphics[width=6 cm]{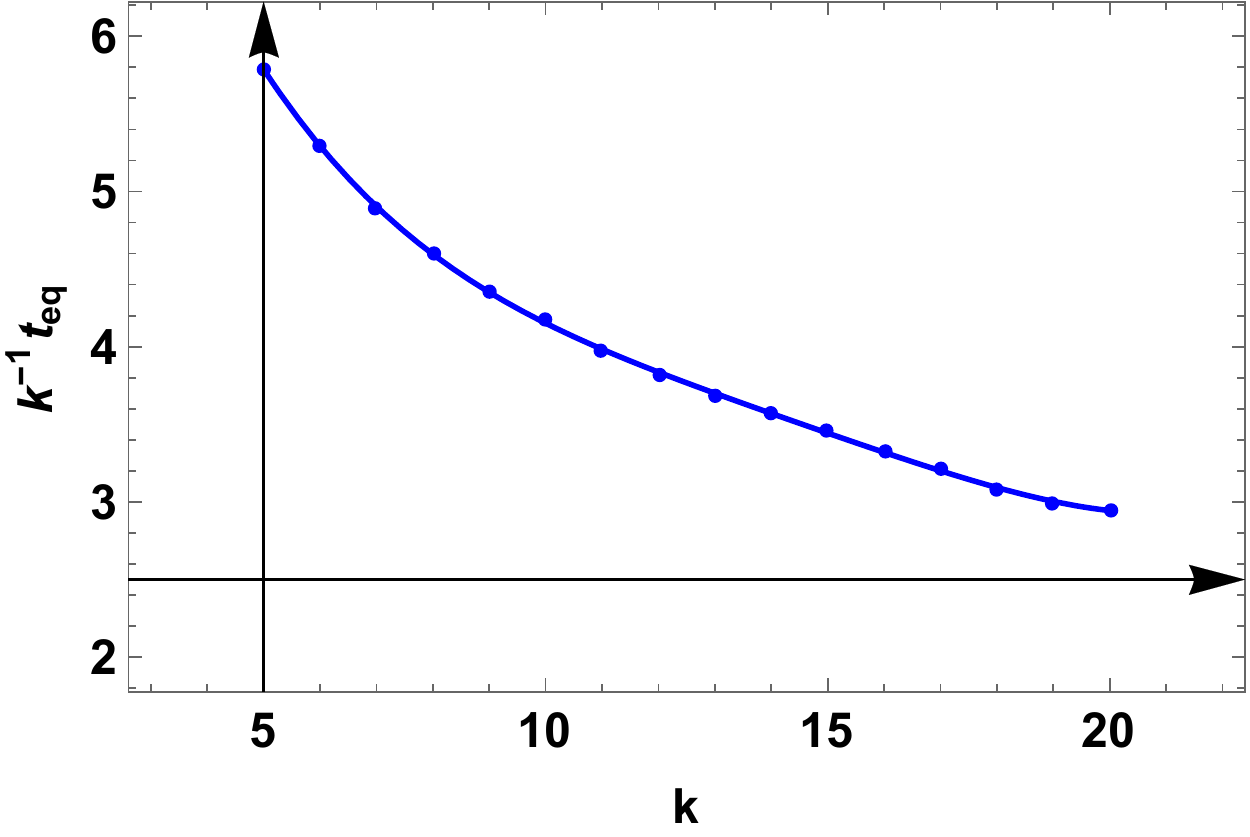}\\
    \includegraphics[width=6 cm]{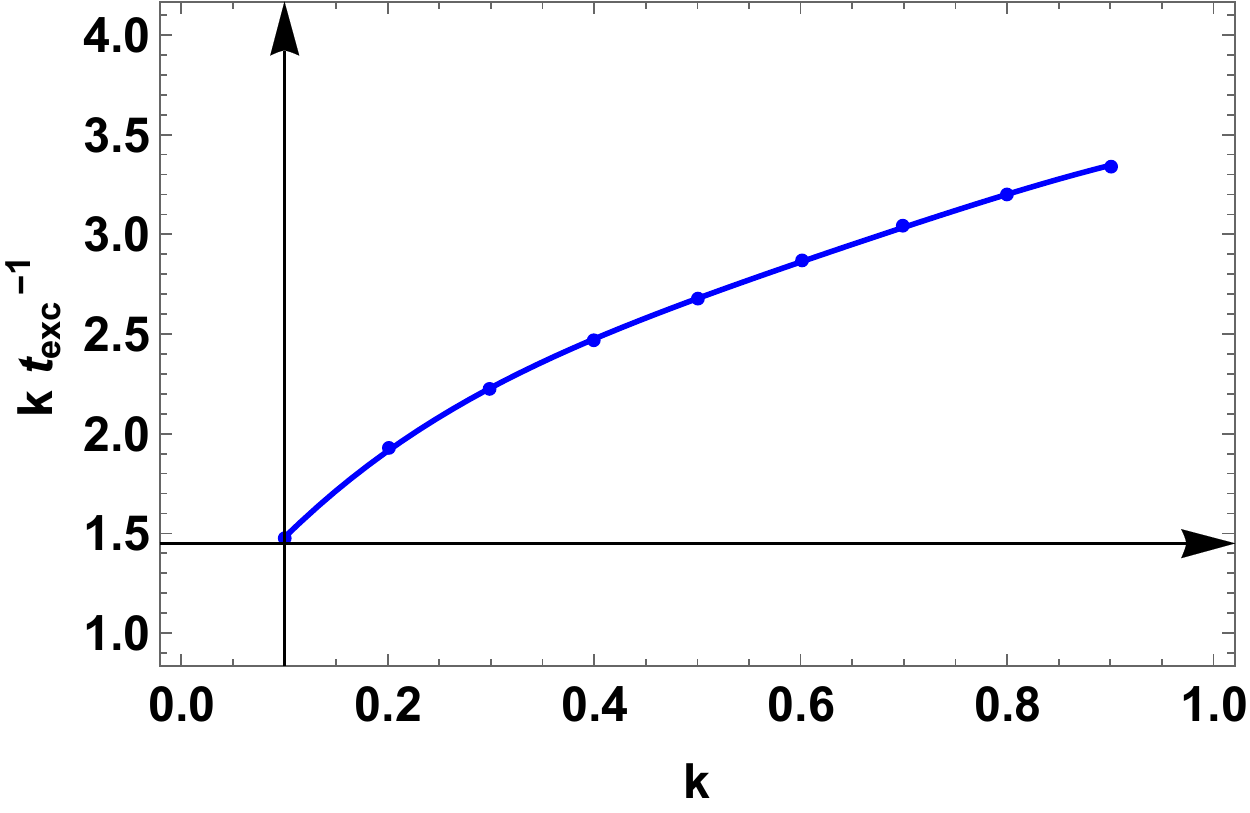}
    \includegraphics[width=6 cm]{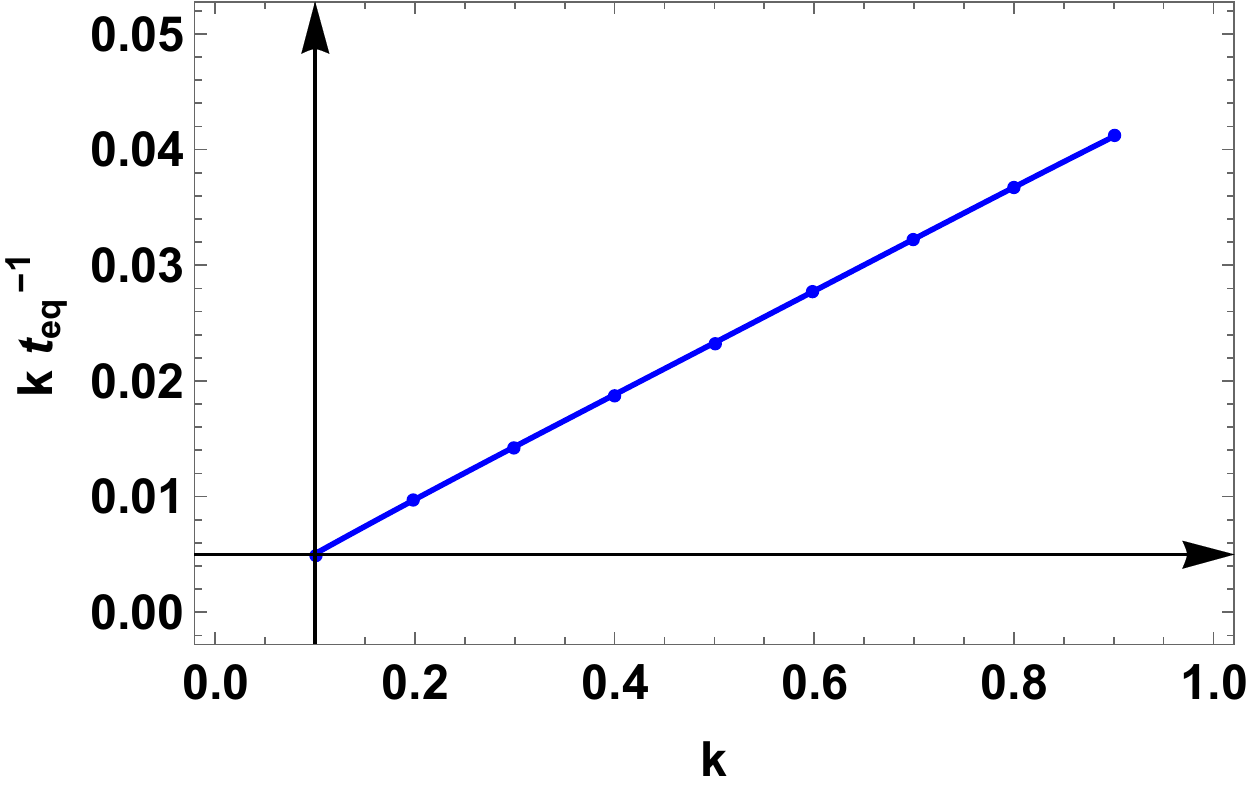}
  \caption{Re-scaled excitation and equilibration time plotted in terms of $k$. For a finite pulse function \eqref{pulsefunctions}, with $k = l$ and $E_0 = 10^{-3}$. The left (right) panel shows the re-scaled excitation time and its inverse (re-scaled equilibration time and its inverse), for large and small transition time. Other types of quenches produce similar results and therefor are not reported here.
}{\label{pulsextra}}
\end{center}
\end{figure}

As it is clearly seen from this figure, for fast quenches, similar to the equilibration time, the inverse of the re-scaled excitation time shows a universal behavior meaning that, for enough small $k$, the value of the $kt_{exc}^{-1}$ is independent of $E_0$. Generally speaking, when investigating either excitation or equilibration, we observe that in fast quenches $k$ is the dominant factor. However, in the opposite limit, by increasing $k$ the final value of the electric field $E_0$ plays more important role and in fact the behavior of the $k^{-1}t_{exc}$ is governed by $E_0$. The same procedure for pulses results in figure \ref{pulsextra} where the behavior of $t_{eq}$ and $t_{exc}$ versus small and large values of $k$, is illustrated..

The other quench types basically produce the same results. 


\section{Apparent Horizon Formation}
\begin{figure}
\begin{center}
    \includegraphics[width=6 cm]{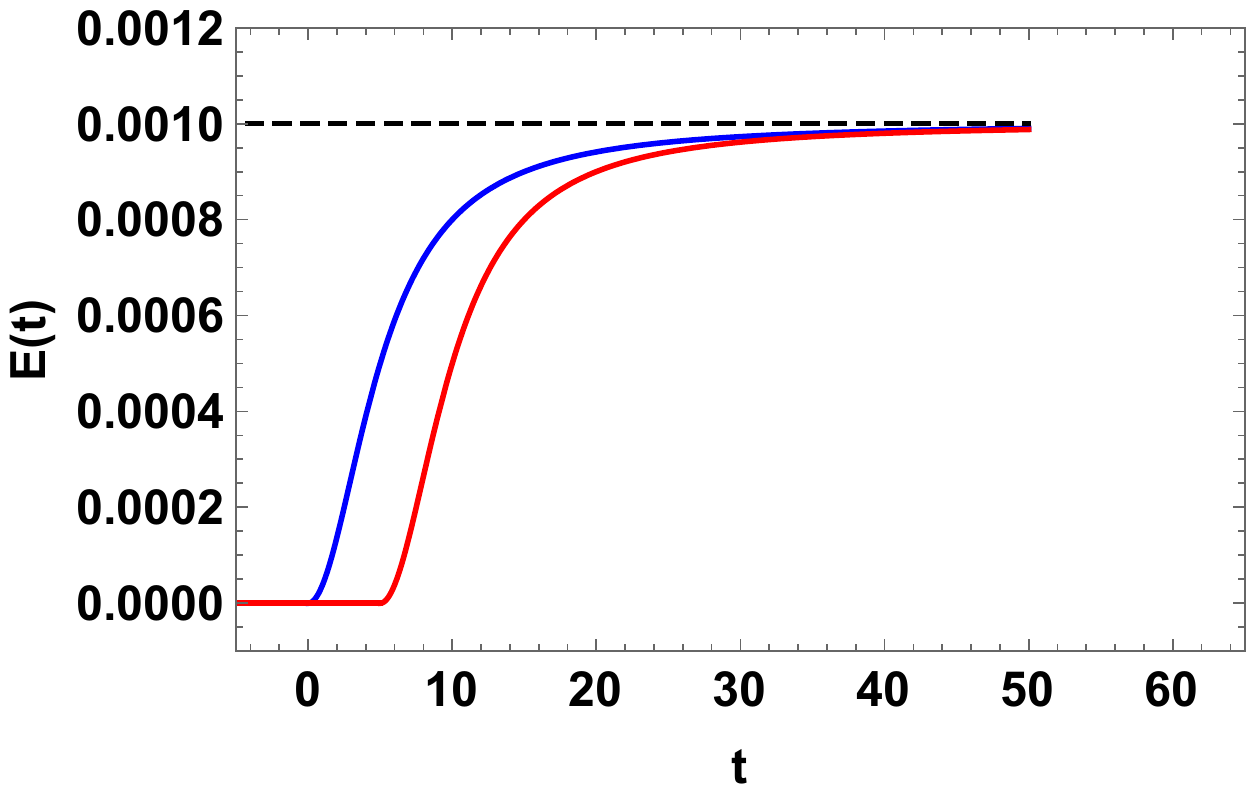}
    \includegraphics[width=6 cm]{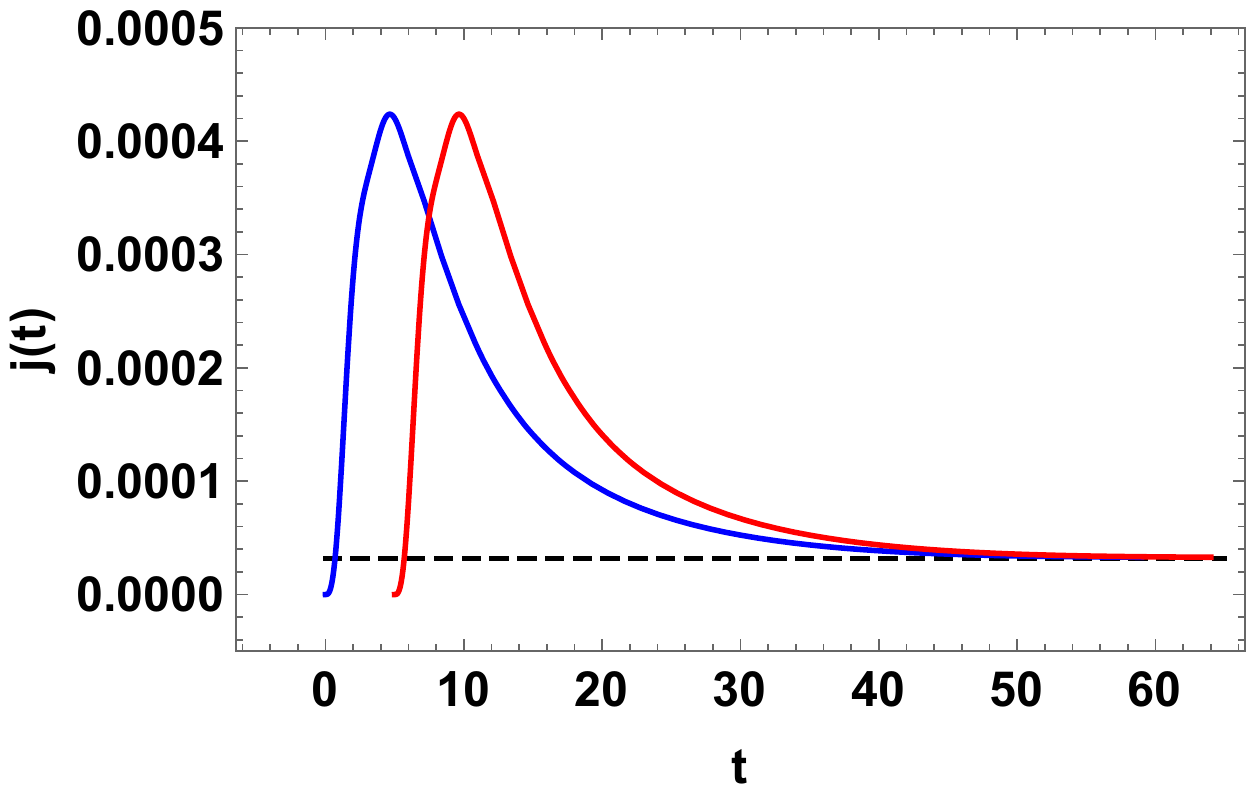}
    \includegraphics[width=8 cm]{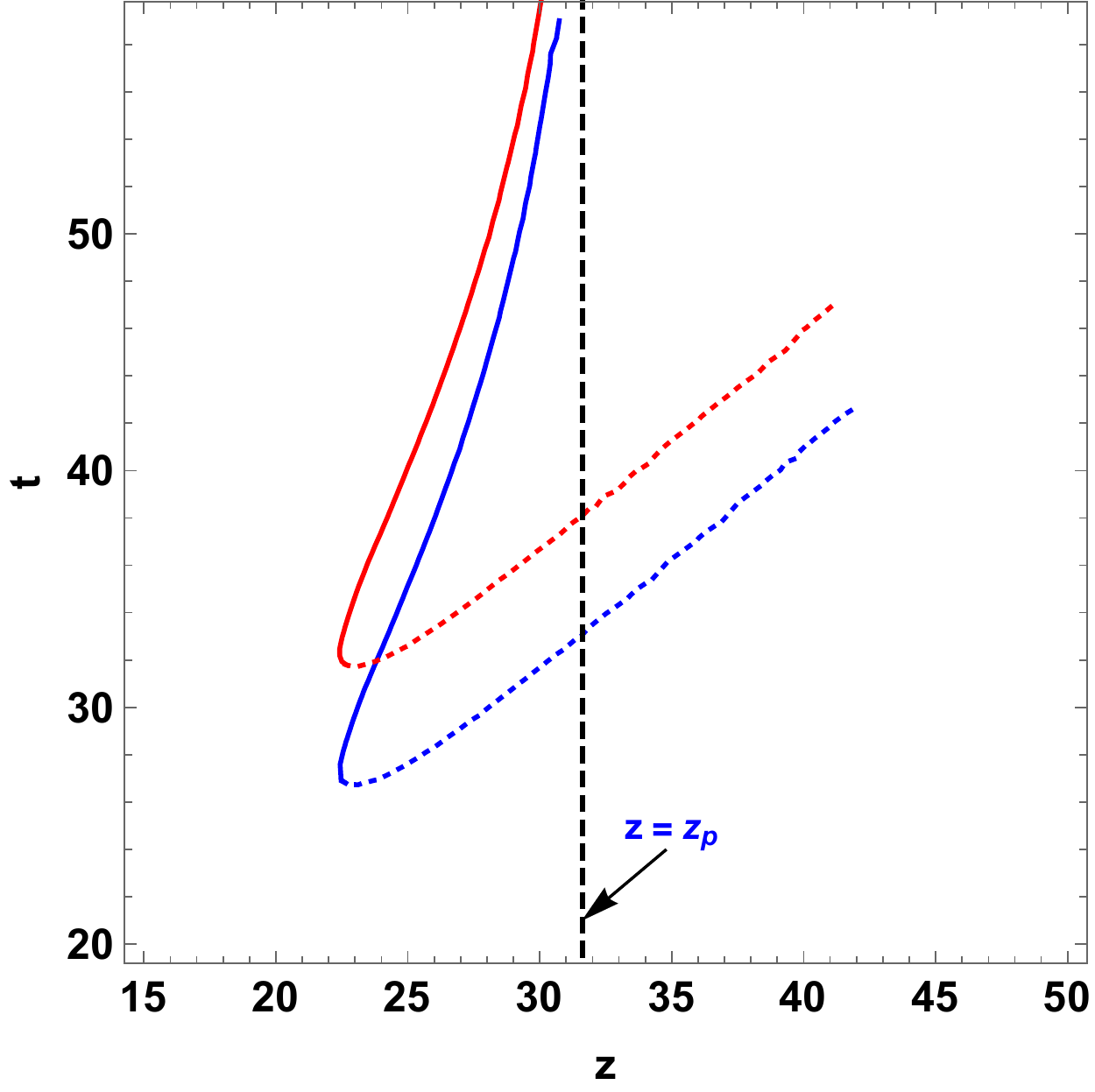}
     \caption{Apparent horizon formation on the probe brane. The conspicuous shift in the tip of the apparent horizon curve (bottom) is due to a time shift in the electric field (top left, half-finite type \eqref{half-finite} with $n = 2$) drawn for $E_0 = 10^{-3}$, $k = 10$ and $\hat{t}=5$. The dotted parts of the apparent horizon curves are not physical.}{\label{delay}}
\end{center}
\end{figure}

In \cite{Hashimoto:2013mua, AliAkbari:2012hb} apparent horizon formation on the probe brane has been studied indicating that a temperature is recognized by the fundamental matter in the gauge theory, at late times. In particular, it is shown that due to energy injection into the system (via a time-dependent electric field like \eqref{finite1} as in \cite{Hashimoto:2013mua}) an apparent horizon is formed on the probe brane. Knowing the time-dependent location of the apparent horizon, $z_{AH}(t)$ as introduced in \eqref{ah}, it is convenient to define an effective temperature 
\be\label{effectivetemperature} %
 T_{eff}=\sqrt{\frac{3}{8\pi^2}} \frac{1}{z_{AH}},
\ee %
and this temperature then approaches the temperature defined in \eqref{temperature static} in the static case, i.e. $z_{AH}(t)\rightarrow z_p$, for $t \rightarrow \infty$, as one can see in figure \ref{delay}. This figure shows that the apparent horizon is formed at the late time of the energy injection, but before that, since the system is out of equilibrium, the apparent horizon does not exist. 

\begin{figure}
\begin{center}
    \includegraphics[width=10 cm]{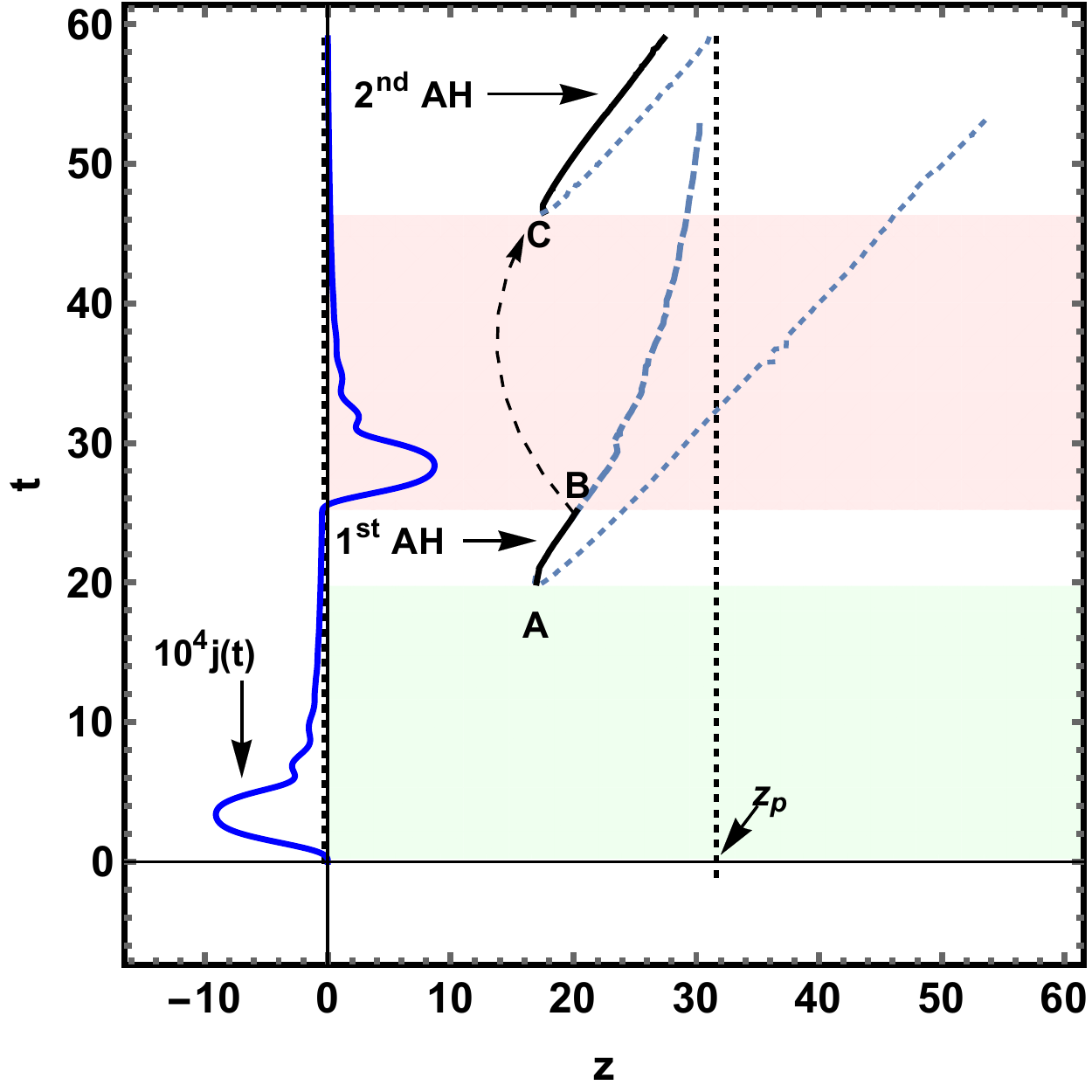}
    \caption{A juxtaposition of the time-dependent current $j(t)$ and the apparent horizon formation. The pulse function considered here is \eqref{pulsefunctions} with $E_0=0.001$, $k=5$ and $l=20$. After switching on an electric field at $t=0$, we are dealing with a non-equilibrium process till $t \sim 20$ (shaded region in the bottom). For $20\lesssim t \leq 25$, the system is near thermal equilibrium and an apparent horizon exists for this duration (the solid curve connecting points $A$ and $B$). At $t= 25$, the electric field starts falling from $E_0$ to zero and hence the system experiences the second non-equilibrium part from $t=25$ till $t\sim 46$ (shaded region in the top). At $t\sim 46$, the second apparent horizon is formed (point $C$). This point marks the beginning of the second part of the thermal path, which drags on forever. In the second part, the system feels a time dependent effective temperature that approaches zero.}{\label{dhorizons}}
\end{center}
\end{figure}

Now let us consider a simple time translation $t \rightarrow t-\hat{t}$ for a quench and see how the apparent horizon is accordingly formed. In fact for $\hat{t}>0\ (\hat{t}<0)$ the time translation for a quench function such as \eqref{finite1} delays (advances) the increase of the electric field. The result of the apparent horizon formation is plotted in figure \ref{delay} which shows that a simple time translation in the quench function would obviously delay (advance) apparent horizon formation, without changing the final temperature, as $z_p$ depends only on the final electric field value, $E_0$. The importance of this shift in the tip of the apparent horizon curve lies in the fact that in our analysis of pulse functions, especially with a rather long and flat top, we split the pulse into a standard quench plus and a shifted standard reverse quench. Each one of these quenches, given enough time, produce their own apparent horizon, and the shift separates them on the $zt$-plane along the $t$-axis (c.f. figures \ref{dhorizons}, \ref{reverseAH}, \ref{doubleriseah}).

The other point that we would like to emphasize on is that the slope of the apparent formation curve in $zt$-plane in figure \ref{delay} is positive and goes to infinity for large $t$. In other words, $z_{AH}(t)$ has an upper limit which is denoted by $z_p$ and marks the formation of an eventual event horizon on the probe brane.

\begin{figure}
\begin{center}
    \includegraphics[width=8 cm]{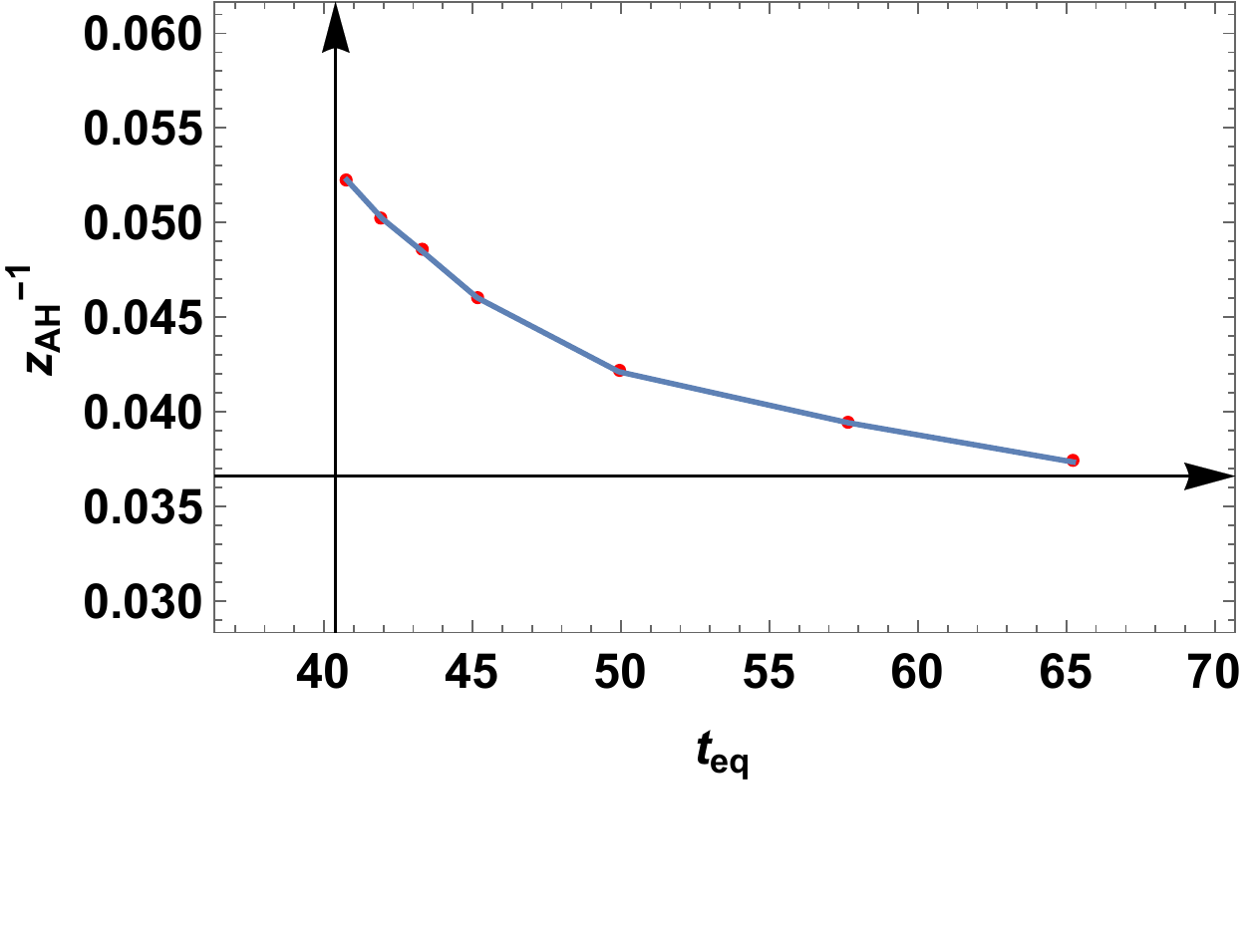}
      \caption{Equilibration time $t_{eq}$ vs. $z_{AH}^{-1}$, drawn for $E_0 = 10^{-3}$ and $k = 5 - 25$, and for a half-finite quench type with $n = 2$}{\label{teq-zah}}
\end{center}
\end{figure}

Due to the fact that the formation of apparent horizon awaits the relaxation of the system, there should be a  connection between the location of the appearance of apparent horizon, $z_{AH}$ (or its corresponding temperature $T_{eff}$), and the equilibration time $t_{eq}$. This is indeed true as one can observe in figure \ref{teq-zah}, which establishes a smooth monotonic dependence between these two important factors of the system. Although the definition of $t_{eq}$ in \cite{Ali-Akbari:2015gba} and this paper depends on statistical criteria (\%5 relative accuracy), but this dependency is not critical to the shape of the $t_{eq} - z_{AH}^{-1}$ curve and a change in the definition (say $\%5 \rightarrow \%10$) would only shift the curve a bit downward as the equilibration time would decrease by this change of standard.

The most interesting situation is when a pulse with a rather large $k$ (and also $l$ for half-finite type like \eqref{pulsefunctions}) is applied. In this case two apparent horizons take form: one due to the rise of the electric field to $E_0$ and the other due to its fall back to zero. In order to investigate this situation we take a look back at the electric current generated by finite type pulses. A review of figure \ref{pulses}, this time specifically for very large $k$ (or $k+l$) as displayed in figure \ref{dhorizons}, reveals that when the electric field remains stabilized for a rather long time, the current has enough time to relax, before eventually getting disturbed by the downfall of the electric field. The conformity of $t_{eq}$ and $z_{AH}^{-1} \propto T_{eff}$ (say, for the first apparent horizon), indicates that an apparent horizon is formed, at least for a while, before the exertion of the second shock (second half of the pulse) which throws the system into a non-equilibrium phase resulting in the loss of apparent horizon.

One surprising feature of figure \ref{dhorizons} is that the highly non-linear equation of motion \eqref{eom} seemingly behaves like a linear partial differential equation, as the two "halves" of the electric field generate their own currents without visibly perturbing each other! And interestingly, it's exactly the second half that causes the second apparent horizon. 
This effect is basically due to the fact that the systems under consideration would not differentiate between an initial constant electric field and an "almost" constant (equilibrated) one.
Mathematically speaking, this shows that the system, considered
as a "dynamical system" in the mathematical terminology, is not
too sensitive to the initial-boundary conditions, and a "chaotic"
behavior is not expected from this type of systems. In other
words, a slight change in the initial-boundary conditions, would
only bring about a slight change in the output.
Physically speaking, one can state this behavior proves that, after a long enough period of equilibration, the system does not remember its past and a new wave of disturbance would act
alone in affecting the system. This phenomenon is also clearly observed in the "double-rise" scenario, in figure \ref{doublerise}. The above result, although applicable to linear systems, is by no means a general feature of non-linear systems.

A close examination also reveals that the first apparent horizon is in accordance with the apparent horizon formation in the quench scenario. It takes form at $t_{eq} \sim 20$, at point $A$, and indeed asymptotically approaches the same $z_p = E_0 ^{-1/2}$ for large enough $t$. But, once the downfall of the electric field is triggered, at point $B$, (at exactly $t = k+l $), the system loses its sense of temperature. Soon, after the second relaxation time kicks in (at $t'_{eq} \sim 46$), the next apparent horizon is formed, and system follows this new thermal path, with its corresponding time-dependent temperature. As one can clearly see, after the point $C$, the second apparent horizon curves upward which removes the possibility of any vertical asymptote, thus resulting in a final zero temperature at the late time. 

\section{Reverse Quench}
This type of quench is basically the same as aforementioned ones only seen in the reverse time order. The electric field is initially around a maximum value of $E_0$ (at the far past), and then reaching $t\sim0$ it starts to relax to its final value of zero (at the far future). The transition from $E_0$ and to zero, could take place in a finite or an infinite process.
\begin{figure}
\begin{center}
    \includegraphics[width=6.5 cm]{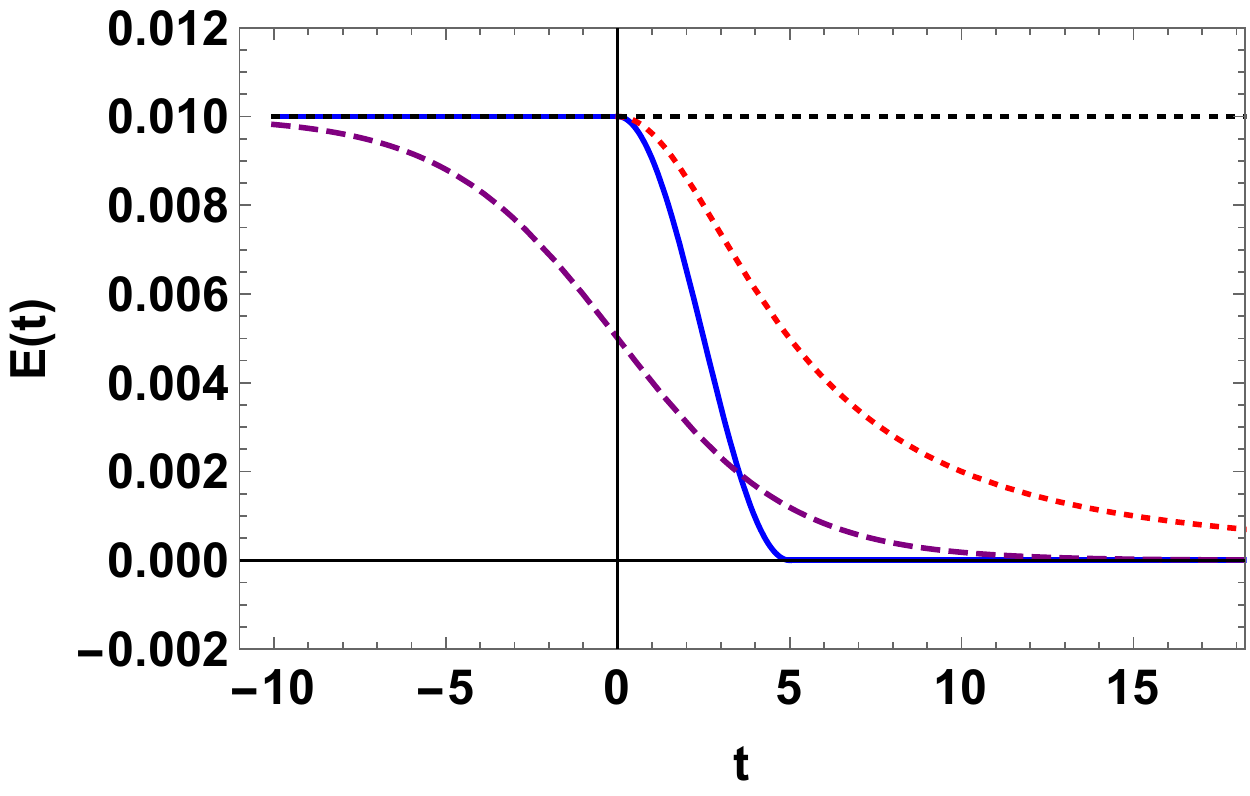}
    \includegraphics[width=6.5 cm]{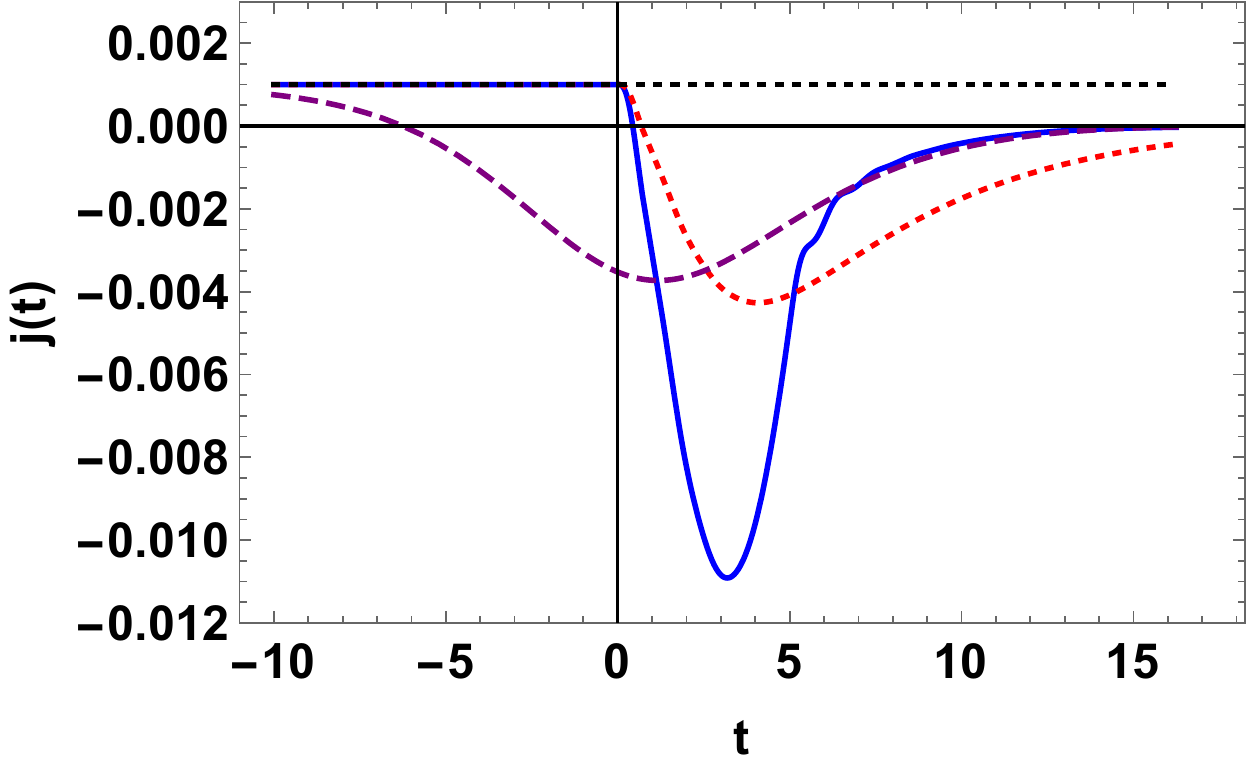}
  \caption{Standard reverse quenches (left), and their resulting currents (right), for $E_0=10^{-2}$ and $k=5$.}{\label{reversequenches}}
\end{center}
\end{figure}
Applying a reverse quench to the system is basically the same as studying the effect of applying a standard quench to the system under time reversal. In this scenario as one would expect, an equilibrated current turns non-equilibrium at $t\sim0$ (the time when the electric field goes through considerable change) and then relaxes to its final value of $j_0=0$, for the late time ($t\longrightarrow+\infty$).

\begin{figure}
\begin{center}
    \includegraphics[width=6.5 cm]{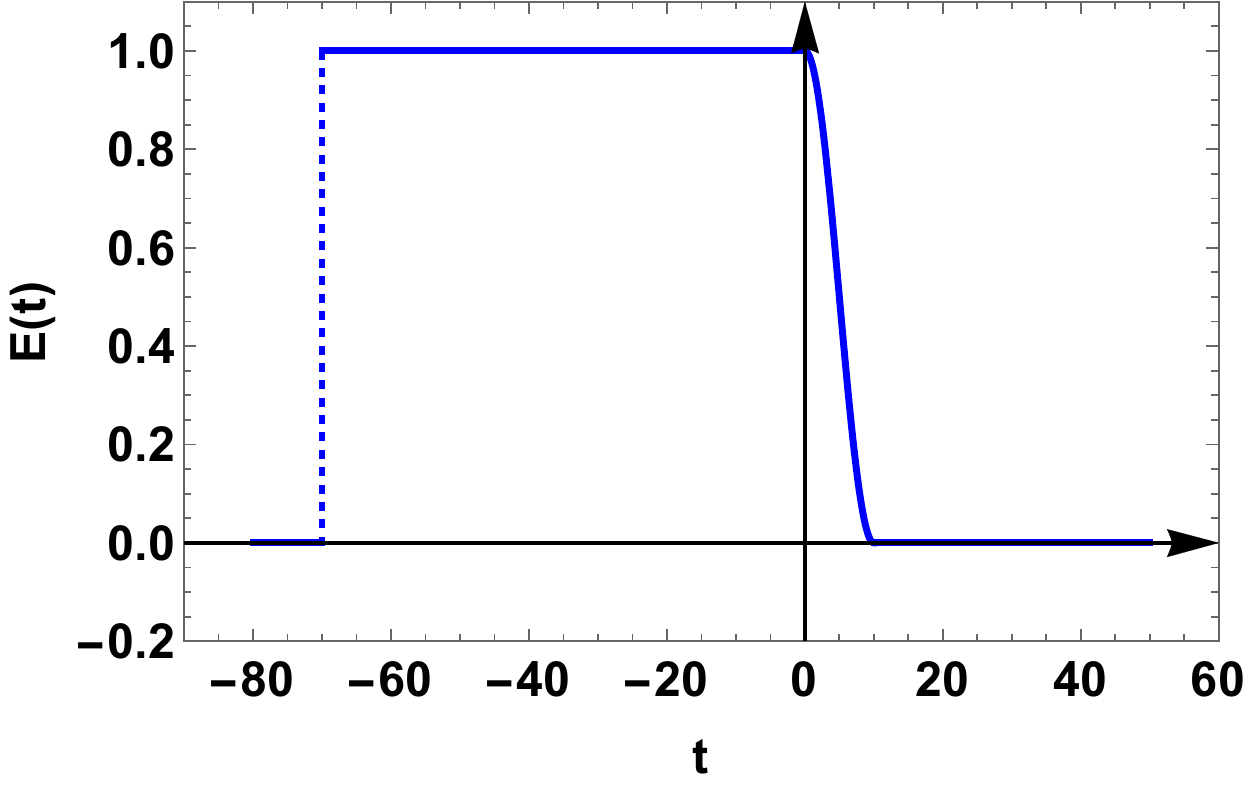}
    \includegraphics[width=6.5 cm]{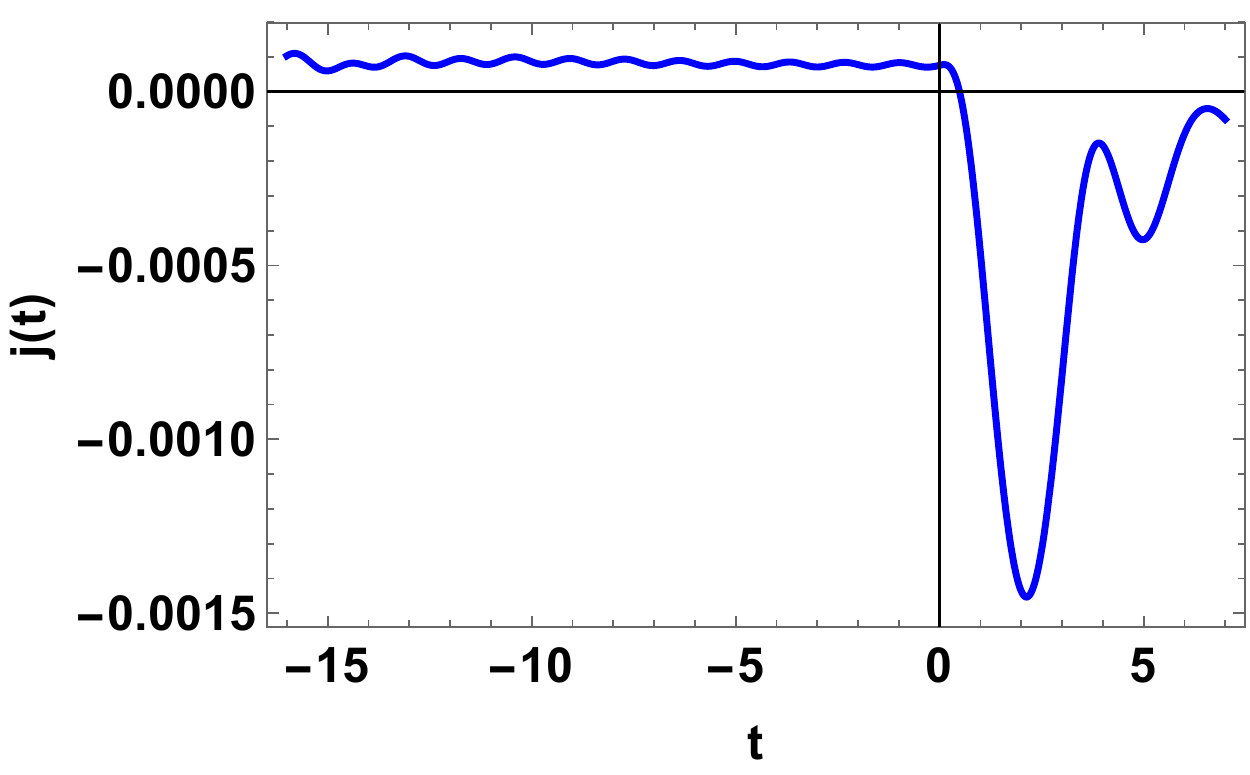}
  \caption{Standard finite type reverse quench and its output current (drawn for $E_0=10^{-3}$ and $k=3$).}{\label{reversenumerical}}
\end{center}
\end{figure}

The standard reverse quenches and their resulting currents are shown in figure \ref{reversequenches}. These diagrams are made  from the diagrams for the standard quenches using time reversal, meaning that all the diagrams are drawn for the same functions except the argument $t$ is replaced with $-t$ (plus a time shift). Instead of time reversal picture, the (reverse) currents could also be read through feeding appropriate initial conditions to the main equation of motion in \eqref{eom}. In order to do this numerically, one can apply an electric field of the type in figure \ref{reversenumerical}, left panel. 
As easily seen from \eqref{ansatz}, $A_x$ contains a term with $\int_{t_0=- \infty}^t E(s)ds$ which would produce a sizeable negative infinity for a constant electric field. This numerical limitation, forces one to start up the electric field at a finite time in the (rather long) past and then change it as smoothly as possible after $t = 0$. Such a treatment would result in adding a large but constant numerical value to the boundary conditions for the equation of motion \eqref{eom} as follows\footnote{In \eqref{initial}, $a$ is a large positive number. To avoid numerical complication, one can start $E(t)$ from certain time $t_0$ in the past. This way $a = -t_0$ becomes finite but forces $E(t)$ to vanish before $t_0$ which would justify the conditions in the bottom line in \eqref{initial}.}
\be
\begin{aligned}
& h(t,z\rightarrow 0)= a E_0, \quad \partial_t h(t,z\rightarrow 0)=0\\
& h(t\rightarrow t_0,z)=0, \quad \partial_z h(t\rightarrow t_0 ,z)=0 \label{initial}
\end{aligned}
\ee
where $a = -t_0$ is a large positive constant.

Such an initial-boundary set of conditions generates the current in figure \ref{reversenumerical}, right panel. The ripples on the tail of the current, are a result of the discontinuity at the start of the electric field. Numerically possible, one can extend the current to as far a negative time as possible. The result, approaches the ideal case of figure  \ref{reversequenches}.
In this case, i.e. when treating the situation numerically, the apparent horizon takes form a few moments after the time when the boundary condition is set (at $t_0$) and approaches its asymptote located at $z=z_p$. 

\begin{figure}
\begin{center}
    \includegraphics[width=10 cm]{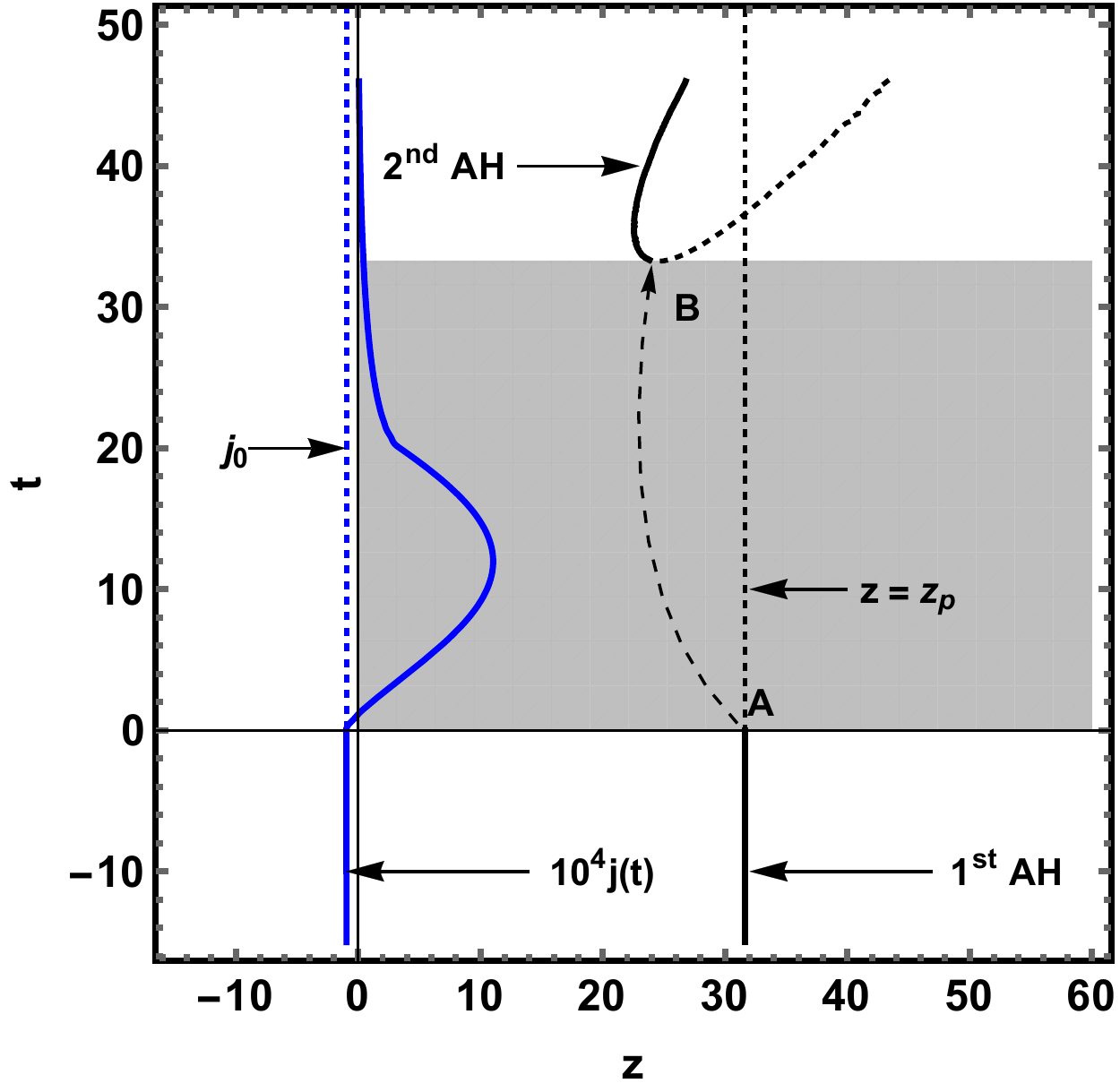}
     \caption{A juxtaposition of a reverse quench of finite type with $E_0 = 10^{-3}$, $k = 20$ and its apparent horizon formation diagram. The system follows the thermal path along the vertical solid line $z = z_p$ (event horizon) from $- \infty$ till $t = 0$, then from the point $A$ to the point $B$ (gray region) the system goes through a non-equilibrium phase. Starting from The point $B$ ($t \sim 34$) the second apparent horizon is formed and the system follows the solid curve from $B$ on.}{\label{reverseAH}}
\end{center}
\end{figure}

Ideally speaking, in this case the apparent horizon formation should happen in the far past and therefore when one moves near $t = 0$, the system should follow the asymptote line, which is indeed the event horizon, as its thermal path. Of course the electric field's sudden change at $t = 0$, renders the rest of this thermal path non-physical and that is when the system awaits the formation of its second apparent horizon to follow it up. The situation is diagrammed in figure \ref{reverseAH}.
The thermal path of the system follows the solid black path, first along the line $z = z_p$ from $-\infty$ to the point $A$, and then from the point $B$ onward. The second part is on a curve that has no vertical asymptote and curves upward (convex), indicating a limiting temperature of zero.

\section{Double-rise Quench}
In this section we briefly review a "double-rise" electric field quench. The results confirm our analysis of double thermalization process discussed in the previous sections.

\begin{figure}
\begin{center}
    \includegraphics[width=6.5 cm]{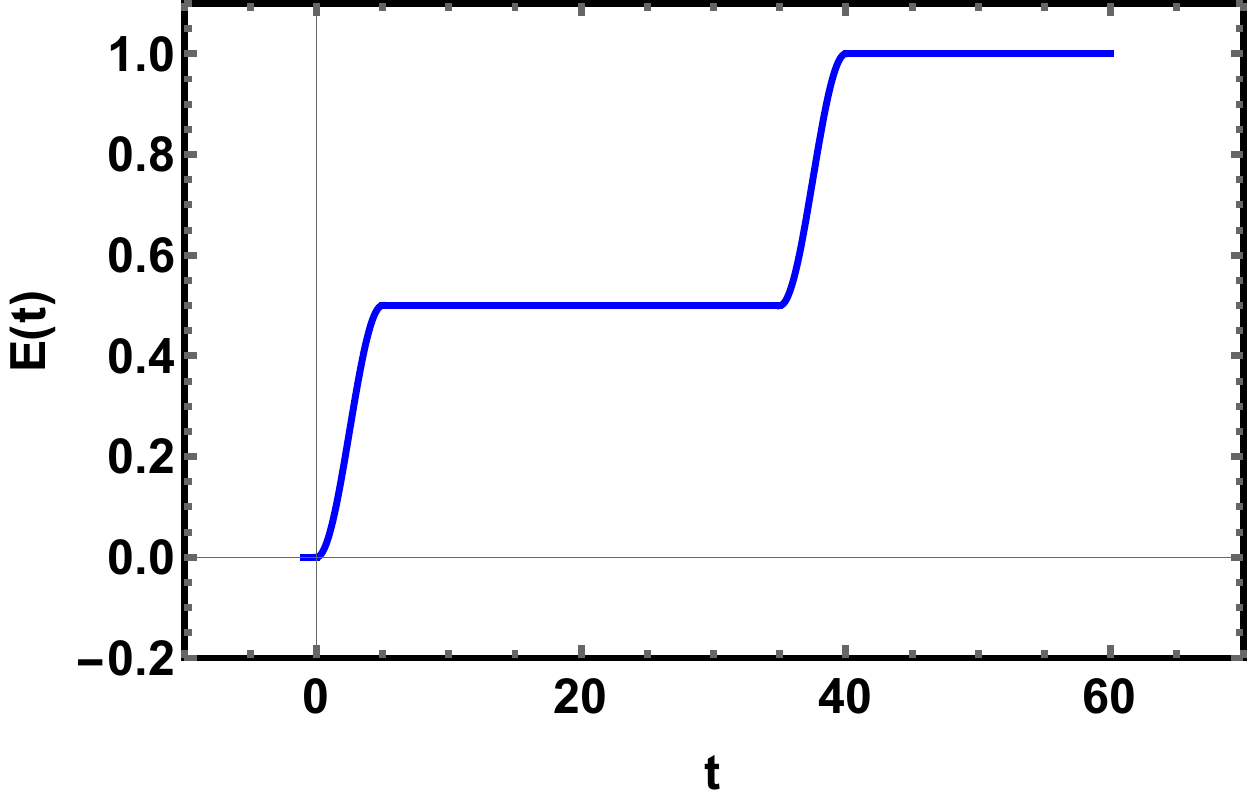}
    \includegraphics[width=6.5 cm]{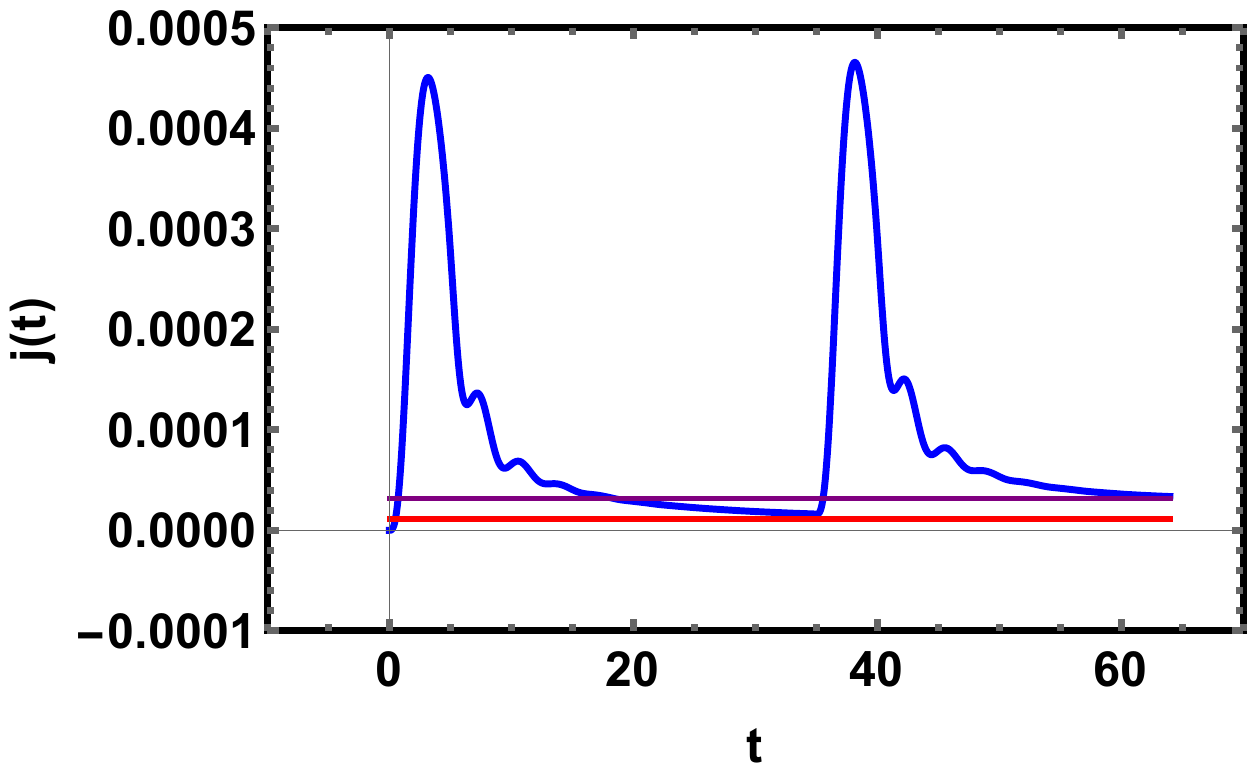}
     \caption{The double-rise quench \eqref{doubleriseeq} and its output current (drawn for $E_0=0.001$, $k=5$, $l=30$). The red and the purple horizontal lines represent stabilized currents $j_1 = (E_0/2)^{3/2}$ and $j_2 = E_0^{3/2}$, respectively.}{\label{doublerise}}
\end{center}
\end{figure}

We subject the system to the following electric field:
\be\label{doubleriseeq}
E(t) = \frac{E_0}{2}
  \begin{cases}
   0 & \text{if } t \leq 0, \\
   \frac{1}{2}(1-\cos(\frac{\pi t}{k}))       & \text{if } 0<t<k, \\
   1       & \text{if } k<t<l+k, \\
\frac{3}{2} + \frac{1}{2} \cos(\frac{\pi(x - l)}{k})  & \text{if } l+k<t<l+2 k, \\
   2 & \text{if } t \geq k,
  \end{cases}
\ee
where, $k$ and $l$, parameters control the shape of field, as seen in figure \ref{doublerise}, left panel.

\begin{figure}
\begin{center}
    \includegraphics[width=10 cm]{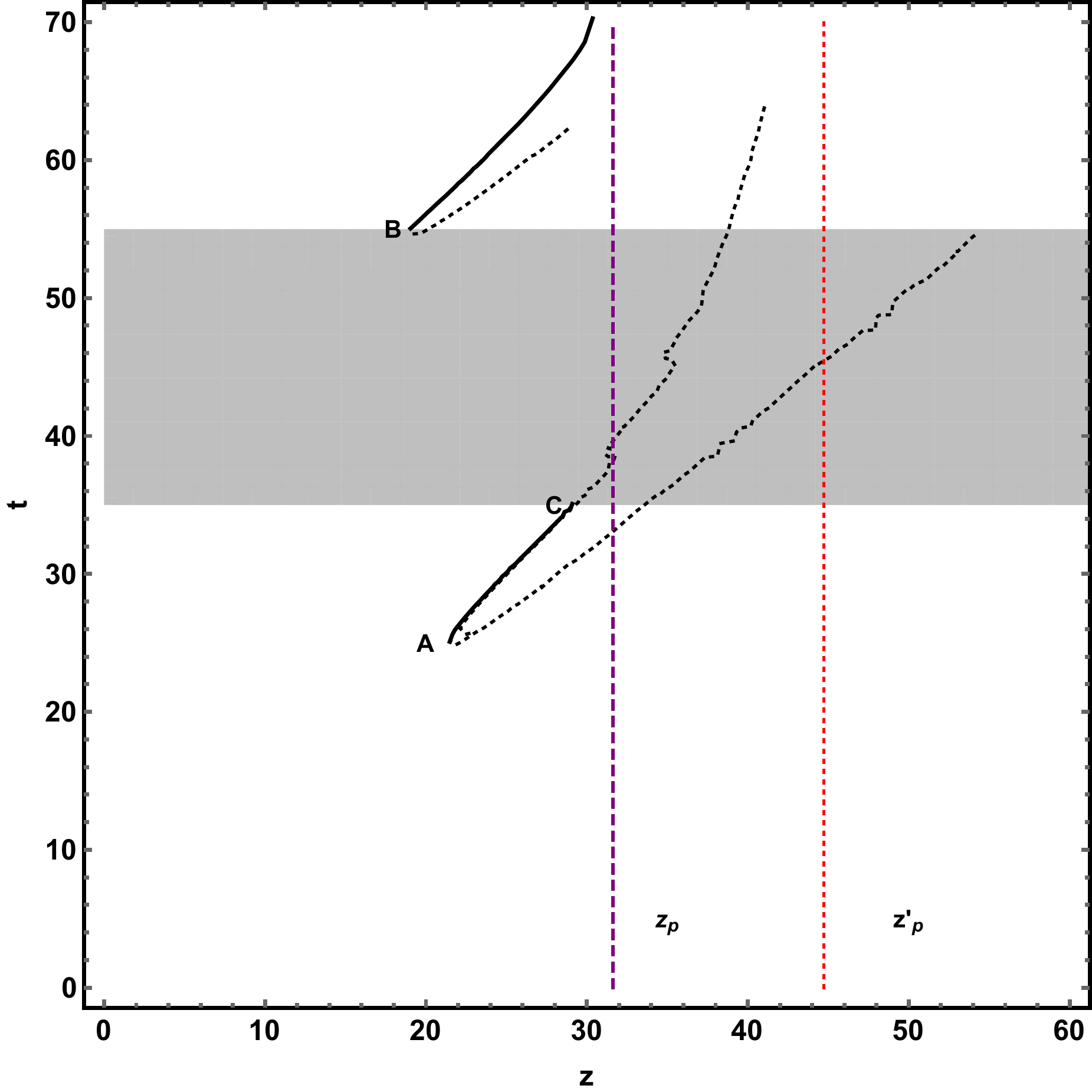}
     \caption{The double apparent horizon formation due to double-rise quench  \eqref{doubleriseeq}. $z'_p = (E_0/2)^{-1/2} $ and $z_p = E_0^{-1/2}$ represent, respectively, the location of the first and the second  event horizons.}{\label{doubleriseah}}
\end{center}
\end{figure}

Physically speaking, this type of quench means raising the electric field in two steps while giving the system enough time in the middle of the process to thermalize. By choosing suitable parameters, the output electric current clearly shows two relaxation periods, one in the middle and one at the end (figure \ref{doublerise}, right panel).
The apparent horizon diagram is in full agreement with the output current and the predicted relaxation periods, as seen in figure \ref{doubleriseah}.

Under the quench \eqref{doubleriseeq}, the system goes through an initial non-equilibrium phase that last till $t \sim 25$, point $A$, when it enters the first relaxation period till $t = 35$, point $C$. During this time-interval the system follows the solid thermal path along the curve $AC$. This curve has an asymptote at $z = z'_p = (0.5 E_0)^{-0.5}$, which represents the temperature corresponding to the first equilibration time. Of course, the second electric field rise would throw the system off of this path and therefore the rest of the first apparent horizon curve after point $C$ ($t > 35$) is non-physical.

At the point $C$, ($t = 35$), the system enters the second non-equilibrium period that lasts till $t \sim 55$, point $B$. From this point on, the second relaxation period begins and the system follows the solid curve starting from $B$ onward. This path that corresponds to the second thermalization, continues for the late time ($t>t_B\sim55$) and during this phase, the system's effective temperature approaches a final value corresponding to $z = z_p = (E_0)^{-0.5}$. The curvatures of both parts of the thermal curve are in agreement with our analysis.
\section*{Acknowledgment}
The authors would like to express their deep gratitude to the Referee for his/her comments that helped making a major improvement in the presentation of the paper.

\end{document}